\documentclass[11pt,onecolumn,draftcls]{IEEEtran}
\usepackage[noadjust]{cite}
\ifCLASSINFOpdf
\else
\fi
\usepackage{mathrsfs, bm, amsthm,caption,subcaption}
\usepackage{amsfonts, amsmath, amsthm, amssymb}
\usepackage{graphicx,color,psfrag}

\newcommand{\as}{\ba}

\newcommand{\ttheta}{\tiltheta} 
\newcommand{\ts}{\tls}

\newcommand{\tp}{\tlp}

\newtheoremstyle{mytheoremstyle} 
    {\topsep}                    
    {\topsep}                    
    {\normalfont}                   
    {}                           
    {\itshape}                   
    {.}                          
    {.5em}                       
    {}  

\theoremstyle{mytheoremstyle}

\newtheorem{fact}{Lemma}
\begin{document}
\bstctlcite{bib:etal}

\newcommand{\bGamma}{\bm{\Gamma}}
\newcommand{\bSigma}{\bm{\Sigma}}
\newcommand{\bOmega}{\bm{\Omega}}
\newcommand{\bPhi}{\bm{\Phi}}
\newcommand{\bPsi}{\bm{\Psi}}
\newcommand{\bLambda}{\bm{\Lambda}}
\newcommand{\bUpsilon}{\bm{\Upsilon}}
\newcommand{\bDelta}{\bm{\Delta}}

\newcommand{\bsigma}{\bm{\sigma}} 
\newcommand{\balpha}{\bm{\alpha}} 
\newcommand{\bphi}{\bm{\phi}}
\newcommand{\bpi}{\bm{\pi}}
\newcommand{\bdelta}{\bm{\delta}}
\newcommand{\bomega}{\bm{\omega}}
\newcommand{\bgamma}{\bm{\gamma}}
\newcommand{\bepsilon}{\bm{\epsilon}}
\newcommand{\blambda}{\bm{\lambda}}
\newcommand{\btheta}{\bm{\theta}}
\newcommand{\bpsi}{\bm{\psi}}
\newcommand{\bmeta}{\bm{\eta}}
\newcommand{\bnu}{\bm{\nu}}
\newcommand{\bxi}{\bm{\xi}}
\newcommand{\bzeta}{\bm{\zeta}}

\newcommand{\tilDelta}{\tilde{\Delta}}
\newcommand{\tilTheta}{\tilde{\Theta}}

\newcommand{\tilbeta}{\tilde{\beta}}
\newcommand{\tiltheta}{\tilde{\theta}}
\newcommand{\tilepsilon}{\tilde{\epsilon}}

\newcommand{\bA}{\mathbf{A}}
\newcommand{\bB}{\mathbf{B}}
\newcommand{\bC}{\mathbf{C}}
\newcommand{\bD}{\mathbf{D}}
\newcommand{\bF}{\mathbf{F}}
\newcommand{\bG}{\mathbf{G}}
\newcommand{\bH}{\mathbf{H}}
\newcommand{\bI}{\mathbf{I}}
\newcommand{\bJ}{\mathbf{J}}
\newcommand{\bM}{\mathbf{M}}
\newcommand{\bN}{\mathbf{N}}
\newcommand{\bP}{\mathbf{P}}
\newcommand{\bQ}{\mathbf{Q}}
\newcommand{\bR}{\mathbf{R}}
\newcommand{\bS}{\mathbf{S}}
\newcommand{\bT}{\mathbf{T}}
\newcommand{\bU}{\mathbf{U}}
\newcommand{\bV}{\mathbf{V}}
\newcommand{\bW}{\mathbf{W}}
\newcommand{\bX}{\mathbf{X}}
\newcommand{\bZ}{\mathbf{Z}}

\newcommand{\ba}{\mathbf{a}}
\newcommand{\bb}{\mathbf{b}}
\newcommand{\bd}{\mathbf{d}}
\newcommand{\be}{\mathbf{e}}
\newcommand{\mbf}{\mathbf{f}}
\newcommand{\bg}{\mathbf{g}}
\newcommand{\bh}{\mathbf{h}}
\newcommand{\bn}{\mathbf{n}}
\newcommand{\bp}{\mathbf{p}}
\newcommand{\bq}{\mathbf{q}}
\newcommand{\br}{\mathbf{r}}
\newcommand{\bs}{\mathbf{s}}
\newcommand{\bu}{\mathbf{u}}
\newcommand{\bv}{\mathbf{v}}
\newcommand{\bw}{\mathbf{w}}
\newcommand{\bx}{\mathbf{x}}
\newcommand{\by}{\mathbf{y}}
\newcommand{\bz}{\mathbf{z}}

\newcommand{\hbeta}{\hat{\beta}}
\newcommand{\htheta}{\hat{\theta}}
\newcommand{\hsigma}{\hat{\sigma}}

\newcommand{\hp}{\hat{p}}
\newcommand{\hn}{\hat{n}}
\newcommand{\hr}{\hat{r}}
\newcommand{\hs}{\hat{s}}
\newcommand{\hx}{\hat{x}}

\newcommand{\hN}{\hat{N}}
\newcommand{\hS}{\hat{S}}

\newcommand{\hbSigma}{\hat{\bm{\Sigma}}}

\newcommand{\hba}{\hat{\mathbf{a}}}
\newcommand{\hbs}{\hat{\mathbf{s}}}
\newcommand{\hbu}{\hat{\mathbf{u}}}
\newcommand{\hbv}{\hat{\mathbf{v}}}

\newcommand{\hbN}{\hat{\mathbf{N}}}
\newcommand{\hbP}{\hat{\mathbf{P}}}
\newcommand{\hbR}{\hat{\mathbf{R}}}
\newcommand{\hbS}{\hat{\mathbf{S}}}
\newcommand{\hbW}{\hat{\mathbf{W}}}

\newcommand{\hbtheta}{\hat{\btheta}}

\newcommand{\dif}{\text{d}}

\newcommand{\bbC}{\mathbb{C}}
\newcommand{\bbR}{\mathbb{R}}
\newcommand{\bbN}{\mathbb{N}}
\newcommand{\bbZ}{\mathbb{Z}}

\newcommand{\calA}{\mathcal{A}}
\newcommand{\calB}{\mathcal{B}}
\newcommand{\calC}{\mathcal{C}}
\newcommand{\calD}{\mathcal{D}}
\newcommand{\calE}{\mathcal{E}}
\newcommand{\calF}{\mathcal{F}}
\newcommand{\calH}{\mathcal{H}}
\newcommand{\calI}{\mathcal{I}}
\newcommand{\calN}{\mathcal{N}}
\newcommand{\calM}{\mathcal{M}}
\newcommand{\calR}{\mathcal{R}}
\newcommand{\calS}{\mathcal{S}}
\newcommand{\calT}{\mathcal{T}}
\newcommand{\calV}{\mathcal{V}}
\newcommand{\calW}{\mathcal{W}}
\newcommand{\calX}{\mathcal{X}}
\newcommand{\calY}{\mathcal{Y}}

\newcommand{\scrA}{\mathscr{A}}
\newcommand{\scrW}{\mathscr{W}}

\newcommand{\tlA}{\tilde{A}}
\newcommand{\tlS}{\tilde{S}}

\newcommand{\tlp}{\tilde{p}}
\newcommand{\tls}{\tilde{s}}
\newcommand{\tlv}{\tilde{v}}

\newcommand{\tlcalA}{\tilde{\calA}}
\newcommand{\tlcalB}{\tilde{\calB}}
\newcommand{\tilcalB}{\tilde{\calB}}
\newcommand{\tlcalI}{\tilde{\calI}}

\newcommand{\tlba}{\tilde{\ba}}
\newcommand{\tlbb}{\tilde{\bb}}
\newcommand{\tlbd}{\tilde{\bd}}

\newcommand{\tlbS}{\tilde{\bS}}

\newcommand{\barn}{\bar{n}}
\newcommand{\barr}{\bar{r}}
\newcommand{\bary}{\bar{y}}

\newcommand{\barS}{\bar{S}}
\newcommand{\barX}{\bar{X}}
\newcommand{\barY}{\bar{Y}}

\newcommand{\barba}{\bar{\ba}}
\newcommand{\barby}{\bar{\by}}
\newcommand{\barbz}{\bar{\bz}}

\newcommand{\tlbA}{\tilde{\bA}}
\newcommand{\tlbR}{\tilde{\bR}}

\newcommand{\tlbp}{\tilde{\bp}}
\newcommand{\tlbs}{\tilde{\bs}}
\newcommand{\tlbv}{\tilde{\bv}}

\newcommand{\tc}{\text{c}}
\newcommand{\td}{{\text{d}}}

\newcommand{\suml}{\sum\limits}
\newcommand{\prodl}{\prod\limits}
\newcommand{\minl}{\min\limits}
\newcommand{\maxl}{\max\limits}
\newcommand{\infl}{\inf\limits}
\newcommand{\supl}{\sup\limits}
\newcommand{\liml}{\lim\limits}
\newcommand{\intl}{\int\limits}
\newcommand{\bigcupl}{\bigcup\limits}

\newcommand{\opconv}{\text{conv}}

\newcommand{\eref}[1]{(\ref{#1})}

\newcommand{\sinc}{\text{sinc}}
\newcommand{\tr}{\text{Tr}}
\newcommand{\var}{\text{Var}}
\newcommand{\cov}{\text{Cov}}
\newcommand{\tth}{\text{th}}

\newenvironment{vect}{\left[\begin{array}{c}}{\end{array}\right]}
\newtheorem{theorem}{Theorem}
\newtheorem{lemma}{Lemma}

\renewcommand{\bs}{\mathbf{d}}
\newtheorem{cor}{Proposition}
\newtheorem{Proposition}{Proposition}
\newtheorem{Definition}{Definition}

\title{Performance Analysis of Parameter Estimation Using LASSO}
%
%
%

\author{Ashkan~Panahi,~\IEEEmembership{Student Member,~IEEE,}
        and Mats~Viberg,~\IEEEmembership{Fellow,~IEEE,}
}
%
%

\markboth{Submitted to Transactions on Signal Processing}%
{Panahi and Viberg }
%



\maketitle
\begin{abstract}
The Least Absolute Shrinkage and Selection Operator (LASSO) has gained attention in a wide class of continuous parametric estimation problems with promising results. It has been a subject of research for more than a decade.  Due to the nature of LASSO, the previous analyses have been non-parametric. This ignores useful information and makes it difficult to compare LASSO to traditional estimators. In particular, the role of the regularization parameter and super-resolution properties of LASSO have not been well-understood yet. The objective of this work is to provide a new insight into this context by introducing LASSO as a parametric technique of a varying order. This provides us theoretical expressions for the LASSO-based estimation error and false alarm rate in the asymptotic case of high SNR and dense grids. For this case, LASSO is compared to maximum likelihood and conventional beamforming. It is found that LASSO loses performance due to the regularization term, but the amount of loss is practically negligible with a proper choice of the regularization parameter. Thus, we provide suggestions on the selection of the regularization parameter. 
Without loss of generality, we present the comparative numerical results in the context of Direction of Arrival (DOA) estimation using a sensor array.
\end{abstract}



%
\IEEEpeerreviewmaketitle
\section{Introduction}
\IEEEPARstart{T}{he} last two decades witnessed rapid emergence of sparse data models and their corresponding techniques in many traditional signal processing areas \cite{genomics, parvaresh2008recovering, EEG, Finance, yao2011compressive}. Although the basic principles of sparsity are easily recognized in many conventional methods, more exotic approaches such as $\ell_1$ penalized least square, well known as LASSO \cite{tibshirani} (Least Absolute Shrinkage and Selection Operator), basis pursuit \cite{basis} or global matched filter \cite{fuchs_GMF}, and its variants \cite{GPSR,AdaptiveLASSO,BLASSO,bayesianLASSO} have been unknown to the majority of the researchers until recently. Regarding these new techniques, it is natural to question how these sparsity-based methods improve the conventional techniques. This is especially important since the new methods demand substantially higher computational effort. In addition, many related questions such as the role of the regularization parameter and the effect of off-grid parameters in LASSO remain unclear. Hence, the current study is devoted to an analysis of LASSO, which provides both a framework to compare LASSO to traditional techniques and a deeper insight into the above questions.

LASSO is a smart solution to the Atomic Decomposition Problem (ADP), first formulated by Chen and Donoho \cite{basis}. Many other techniques such as matching pursuit \cite{MP} and orthogonal matching pursuit \cite{OMP}, Least Angle Regression (LARS) \cite{LARS}, and Compressive Sampling Matching Pursuit (CoSaMP) \cite{CoSaMP} are also developed to solve the ADP. The ADP naturally appears in various problems, e.g. the ones dealing with physical fields \cite{lustig2007sparse,yao2011compressive,provost2009application}. It invokes sparsity, since it may be viewed as a specific type of the so called Union-of-Subspaces (UoS) model, where subspaces are constructed from a set of dictionary bases \cite{Eldar2009}. However, LASSO is only well-defined for a finite dictionary case, while the problems of interest herein are normally related to infinite dictionaries. Examples of such are  frequency and spectrum estimation \cite{Eldar2010,tropp2010beyond}, sensor array analysis \cite{Malioutov}, image processing \cite{baraniuk2007compressive, arigovindan2005variational}, tomography \cite{lustig2007sparse,milanfar1996moment} and seismology \cite{yao2011compressive}. In practice, this is rectified by adopting a dictionary sampling (discretization) scheme, which provides a set of \emph{quantized} estimates \cite{herman2009high}. This is similar in spirit to the so-called spectral based techniques such as matched filter banks \cite{stoica1998matched}, but different in that LASSO provides a strongly sparse spectrum \cite{Malioutov}. Another difficulty arises in selecting the LASSO Regularization Parameter (RP). In essence, this reflects the freedom in selecting the model order. However, this is particularly difficult since the relation between RP and the model order is generally complex \cite{zhao2006model}. 

Many other pioneering works have considered analyzing LASSO, commonly focusing on an information theoretic aspect, widely referred to as compressive sensing (CS) \cite{Donoho_CS}. In other words, LASSO has been studied as a \emph{decoder}, which together with a random linear encoding scheme provides a capacity achieving (optimal) compression rate in asymptotically large setups. However, as we show here, the asymptotic analysis, such as the ones in \cite{candes-near,Donoho_CS,RIPless,baraniuk2007compressive,oymak2013squared} rely on techniques which neglect useful information of LASSO, making them unsuccessful in explaining various effects such as discretization and regularization. Consequently, the final results of a CS study is incompatible with a classical analysis of an estimation problem. The same concern is also observed in some other works, e.g. \cite{candes2012towards, tang2012compressive}. In the above studies, it is observed that the difficulty apears since the sparsity pattern (support) is expressed implicitly (non-parametric). Thus, we suggest to fill the above gap by providing an analysis, where the support is explicitly expressed by parameters (parametric). In the previous literature, one may find similar attempts such as \cite{ben2009cramer}. However, the considered metrics therein are not sufficient for the above mentioned practical interest.  

From a parametric point of view, an ADP is a variable-order problem, where LASSO simultaneously provides an estimator and an order selection scheme. Taking this perspective and similar to many classical studies, we consider an individual case analysis, enabling comparison to the Cramer Rao Bound (CRB) \cite{stoica1990performance}. This also brings a new insight into the problem of RP selection as an order selection technique. 
We also consider an asymptotically high SNR analysis to enjoy linearization techniques. 
We also address the discretization problem. Similar to the spectral-based techniques, our approach is to find an intermediate continuous estimator, of which the LASSO estimates can be regarded as a quantization. In simpler words, we show that the LASSO estimates converge to the intermediate estimates, called Continuous LASSO (CLASS) estimates, when we employ an increasingly dense discretization. The idea of CLASS is rapidly emerging in the ongoing research literature \cite{bhaskar2011atomic,Panahi2014Gridless}. Thus, the current work can also be considered as an analysis of the more recent techniques of solving the ADP.  Clearly, the implementation aspect of CLASS is irrelevant to the current study as it only serves as a bridge to analyze LASSO. The LASSO error is then identified as the combination of the CLASS error and the trivial quantization noise imposed by discretization. 

Employing the above, we obtain the following results. First, we find the explicit relation between error, noise and the RP. This confirms that the RP introduces an undesired bias. However, unlike the Fourier-based techniques, the bias is proportional to the noise and vanishes in the noiseless case. Another important observation is that the behavior of LASSO in the noiseless case is completely independent of the signal power. Note that in presence of sources with high dynamic range, other state of the art techniques such as RELAX \cite{li1996efficient} and SAGE \cite{fessler1994space} are well known to behave poorly. Then, we discuss a certain strategy of RP selection and formulate the overall mean squared error (MSE) corresponding to the selected strategy.  These results generally show that although LASSO does not achieve the CRB due to the regularization induced bias, in many occasions the degradation is negligible.

In summary, the novel ideas and results of this paper are the following:
\begin{itemize}
\item We introduce a framework, enabling to compare LASSO with other parameter estimation techniques.
\item We provide expressions for the LASSO estimation error in our developed framework. 
\item Based on the expressions, we provide some suggestions for the selection of RP. 
\item We compare the resulting expressions to the error of the previously analyzed techniques, namely RELAX and conventional beamforming, as well as the CRB. We conclude that while the LASSO technique is substantially more robust assuming high dynamic range of amplitudes, in many practical situations, it loses a negligible amount of performance due the biasing effect of regularization.
\end{itemize}

\section{Mathematical Modeling}

Consider a closed index set $\Theta\subset\mathbb{R}$ and a collection of complex \emph{basis} vectors $\as(\theta)\in\mathbb{C}^m$ indexed by the elements $\theta\in\Theta$. For our purpose, it suffices to assume that $\as(\theta)$ is a smooth function of $\theta$, where it is referred to as a \emph{manifold}.  In most applications of interest, the dependence of $\as(\theta)$ on $\theta$ is non-linear. Consider a set of $n$ indexes $\btheta=[\theta_1,\theta_2,\ldots,\theta_n]$ and its corresponding discrete-time \emph{complex amplitudes}  $\mathbf{s}(t)=[s_1(t), s_2(t),\ldots,s_n(t)]\in\mathbb{C}^n$ for $t=1,2,\ldots, T$.  We refer to the expression
\begin{equation}\label{p2:eq:premodel}
\mathbf{x}(t)=\sum\limits_{k=1}^n\as(\theta_k)s_k(t)
\end{equation}
as an atomic synthesis. For simplicity we denote a synthesis such as \eqref{p2:eq:premodel} by $\{(\theta_k,\{s_k(t)\})\}$.  In many cases, the synthesized vectors $\mathbf{x}(t)$ correspond to a sequence of observed data and the pair of indexes and amplitudes are to be estimated. This is called an atomic decomposition problem (ADP) \cite{basis}. We call a basis manifold $\ba(\theta)$  \emph{regular} if any arbitrary sequence $\{\mathbf{x}(t)\}$ can be decomposed as in \eqref{p2:eq:premodel}. We assume that $\ba(\theta)$ is regular throughout this paper.  Generally speaking, the order $n$ may or may not be known. In either case, the model in \eqref{p2:eq:premodel} may be insufficient to uniquely infer the decomposition $\{(\theta_k,\{s_k(t)\})\}$ from observations $\{\bx(t)\}$. If $n$ is unknown, the ADP model is commonly accompanied by the principle of parsimony, stating that the smallest order $n$, usually referred to as data rank, is always preferable. In this case, the corresponding synthesis is often referred to as an ideal ADP. Of course, this not generally appealing in presence of noise, which is shortly discussed.  

It is also useful to consider the smallest number $n_0$ of linearly dependent atoms $\{\as(\theta_k)\}$, which is sometimes denoted by Spark$(\as)$ \cite{donoho2003optimally}. In this case, the manifold $\as(\theta)$ is also called $n_0-$ambiguous. Clearly, this is only possible if $n_0\leq m+1$. 
Moreover, one can easily construct an $m+1-$ambiguous manifold in an $m-$ dimensional space. A practical example of such is the Uniform Linear Array (ULA) manifold, discussed in Section \ref{p2:sec:array}. The following simple but fundamental result according to \cite{donoho2003optimally} formulates the uniqueness of the ideal ADP.

\begin{theorem}\label{p2:theorem:fund_uniq}
If a manifold is $n_0-$ambiguous, each set of observations $\{\bx(t)\}$ has at most one ideal decomposition $\{(\theta_k,\{s_k(t)\})\}$ of an order $n< n_0/2$.
\end{theorem}


Another issue with ADP is that the observed data is normally noisy. Thus, it is more practical to assume a model of the following form
\begin{equation}\label{p2:eq:model}
\mathbf{x}(t)=\sum\limits_{k=1}^n\as(\theta_k)s_k(t)+\bn(t)
\end{equation}
where the noise vector $\bn(t)$ is assumed to be a white and circularly symmetric, complex-valued Gaussian process throughout this study. Given $T$ data snapshots $\{\bx(k)\}$,
the problem of interest here is to estimate the signal parameters $\theta$ and their corresponding amplitudes $s$. If the model order is unknown, it also needs to be estimated. The focus in this paper is to asses the quality of the parameter estimates $\hat{\theta}$.

For the noisy model in \eqref{p2:eq:model}, the Least Squares (LS) solution of ADP for a given order $n$ is given by  
\begin{equation}\label{p2:eq:LS}
\min\limits_{\calM_n}\sum\limits_{t=1}^T\left\|\bx(t)-\sum\limits_{k=1}^n\as(\theta_k)s_k(t)\right\|_2^2
\end{equation}
where $\calM_n$ denotes the set of all irreducible decompositions $\{(\theta_k,\{s_k(t)\})\}$ of order $n$. The LS solution in \eqref{p2:eq:LS} coincides with the Maximum Likelihood  (ML) estimator, providing interesting statistical properties \cite{scharf1991statistical,stoica1990performance}.

In ignorance of the order $n$, the previous statement of the principle of parsimony should be utilized with extra care. Note that unlike \eqref{p2:eq:premodel}, the model in \eqref{p2:eq:model} admits any order $n$. However, lower order expressions associate with higher magnitude residual. Thus, a more sophisticated modification of \eqref{p2:eq:LS} should be considered. This is usually referred to as the Model Order Selection (MOS) problem, which essentially establishes a balance between the residual level and the order \cite{stoica2004model,rissanen1983universal}.

\subsection{The Principle of Sparsity}\label{p2:sec:model:sparsity}

Solving the LS problem in \eqref{p2:eq:LS} and MOS has been previously considered. For reviews on different aspects of the problem, see \cite{stoica1989music,ottersten1993exact}, where the accuracy of the estimates for $\{\theta_k\}$ and $\{s_k(t)\}$ are also discussed in different asymptotic cases. Such analysis and any numerical method to solve \eqref{p2:eq:LS} by assigning and iteratively updating values to parameters $\{\theta_k\}$ is commonly called \emph{parametric}.  Despite their theoretical accuracy, the parametric approaches suffer from numerical deficiency, which has motivated for alternative approaches. LASSO is regarded as such a \emph{non-parametric} method, which has its roots in what we refer to as the principle of sparsity, explained below. 

The principle of sparsity simply refers to the fact that in \eqref{p2:eq:premodel}, the parameters corresponding to zero amplitude are ignorable. Note that taking any decomposition $A=\{(\theta_k,\{s_k(t)\})\}$ of an order $n$, one may define $\btheta_s=\{\theta_{k_1},\theta_{k_2},\ldots,\theta_{k_r}\}$, the subset of $\{\theta_k\}$ comprising the elements $\theta_k$ for which $s_k(t)\neq 0$ for at least a single snapshot $t$. Equivalently, $\theta_k\notin\btheta_s$ implies that $s_k(t)=0$ for every time index $t$. The set $\btheta_s$ and its number of elements are called the \emph{support} and \emph{cardinality} of the decomposition, and are denoted by $\text{Supp}(A)$ and $\|A\|_0$, respectively. Then, define the \emph{reduced} decomposition $B=\{(\theta_{k_l},\{s_{k_l}(t)\})\}$. Note that
\begin{equation}\label{p2:eq:sparsity}
\mathbf{x}(t)=\sum\limits_{k=1}^n\as(\theta_k)s_k(t)=
\sum\limits_{\theta_k\in\btheta_s}\as(\theta_k)s_k(t)+
\underbrace{\sum\limits_{\theta_k\notin\btheta_s}\as(\theta_k)s_k(t)}_\mathbf{0}=
\sum\limits_{\theta_k\in\btheta_s}\as(\theta_k)s_k(t)
\end{equation}
Thus, $B$ synthesizes the same vectors $\bx(t)$ as $A$ and the difference between $B$ and $A$ is practically unimportant. We call the reduced decomposition $B$ the \emph{root} of the original decomposition $A$. If a decomposition  $\{(\theta_k,\{s_k(t)\})\}$  is such that for any $k$ the amplitude $s_k(t)$ is nonzero for at least one time index $t$, i.e. it is only reduced to itself, it is called an irreducible decomposition.

A high-order decomposition with a low cardinality is called a sparse decomposition. It can be naturally reduced to a low-order decomposition. Accordingly, LASSO is based on finding a sparse decomposition, which reduces to the ideal decomposition. 
\subsection{The Sensor Array Example}
\label{p2:sec:array}
Finally in this section, we introduce a practical illustrative example which we also consider later. We consider the planar Direction Of Arrival (DOA) estimation problem, in which a set of $m$ sensors listen to $n$ far and narrow band sources and decide on their directions. The received data is modeled by (\ref{p2:eq:premodel}), where the basis manifold is given by (\cite{fuchs_GMF})
\begin{eqnarray}\label{p2:eq:steering}
\as(\theta)=\left[e^{j\frac{2\pi}{d}r_1\cos(\theta-\rho_1)}\ e^{j\frac{2\pi}{d}r_2\cos(\theta-\rho_2)}\ldots\right.\nonumber\\ \left.e^{j\frac{2\pi}{d}r_m\cos(\theta-\rho_m)}\right]^T,
\end{eqnarray}
in which $(r_i,\rho_i)$ is the polar coordinate pair of the $i^{\mathrm{th}}$ sensor $(i=1,2,\ldots,m)$ and $d$ is the wavelength at the central frequency. Then, the goal is to estimate $\{\theta_k\}$ which represents the directions given $\{\bx(t)\}$. Obviously, the problem is defined in a complex-valued space of variables. The manifold in \eqref{p2:eq:steering} is not necessarily unambiguous. An important unambiguous case, which we focus on later is the half-wavelength ($r_i=\frac{(i-1)d}{2}$) Uniform  Linear  Array (ULA). Note that a linear array means that $\rho_i=0$. In this case, it is more convenient to write (\ref{p2:eq:steering}) in terms of the \emph{electrical angle} $\phi=\pi\cos\theta$. The ULA manifold resembles the classical Fourier basis, when represented in terms of the electrical angle. Thus, the sensor array example essentially includes other applications such as frequency estimation and sampling.
 
The ULA manifold is unambiguous. This is easily seen by taking any combination of $m$ distinct bases indexed by electrical angles $\phi_1,\phi_2,\ldots,\phi_m$ and noting that the matrix
\begin{equation}
[\ba(\phi_1)\ \ba(\phi_2)\ldots\ba(\phi_m)]=\left[\begin{array}{cccc}
1& 1& \dots& 1\\
e^{j\phi_1} & e^{j\phi_2} &\dots & e^{j\phi_m}\\
e^{j2\phi_1} & e^{j2\phi_2} & \dots & e^{j2\phi_m}\\
\vdots& \vdots& \ddots& \vdots\\
e^{j(m-1)\phi_1} & e^{j(m-1)\phi_2} & \dots & e^{j(m-1)\phi_m}
\end{array}\right]
\end{equation}
is a Vandermonde matrix and thus it columns are linearly independent as long as they are distinct.

\section{LASSO, Parametric LASSO and CLASS}

In the previous section we formulated atomic decomposition by LS and discussed the principle of sparsity. Sparsity does not directly simplify the computational procedure of obtaining the decomposition. Instead, it provides a framework to obtain better approximate results. For example, greedy algorithms such as Matching Pursuit (MP) and Orthogonal Matching Pursuit (OMP) can be applied due to the principle of sparsity. They basically select bases from $\Theta$ iteratively. Despite their wide application, they have poor theoretical properties. This motivated a different approach by introducing an approximate optimization, whose solution is related to ADP and simple to obtain. A fairly general framework in this matter is to consider convex optimization, where LASSO is a good example. 
To solve the ADP, the LASSO method suggests to consider a finite, but large discretization (grid) $\tilde{\Theta}=\{\ttheta^1,\ttheta^2,\ldots,\ttheta^N\}$ of $\Theta$ and assign parameters $\{\ts^k(t)\}$ to $\ttheta^k$. The parameters $\ttheta^k$ are known. However, if the decomposition $\{\ttheta^k,\{\ts^k(t)\}\}$ is sparse it can be reduced to a good approximation of any desired decomposition. To ensure sparsity, LASSO considers the following optimization
\begin{equation}\label{p2:eq:LASSO}
\min\limits_{\{\ts_k(t)\}}\frac{1}{2}\sum\limits_{t=1}^T\left\|\bx(t)-\sum\limits_{k=1}^N\as(\ttheta_k)\ts_k(t)\right\|_2^2
+\lambda\|\{\ts_k(t)\}\|_{2,1}
\end{equation}
where
\begin{equation}
\|\{\ts_k(t)\}\|_{2,1}=\sum\limits_{k=1}^N\sqrt{\sum\limits_{t=1}^T|\ts_k(t)|^2}
\end{equation}
This is sometimes called group-LASSO to emphasise on its multi-snapshot nature. 
The regularization parameter $\lambda>0$ controls the cardinality of the estimate while the order is fixed to $N$ (see Section \ref{p2:sec:model:sparsity}). However, the analytical relation between $\lambda$ and the cardinality is difficult to obtain. It is also easy to show that group-LASSO always has a solution of a bounded cardinality independently of the order $N$ and different grid choices. For a single snapshot case, $T=1$, group-LASSO is simplified to the more familiar LASSO optimization. We usually apply the term LASSO to also refer to group-LASSO when there is no risk of confusion. Using Lagrangian duality and noting that group-LASSO is a convex optimization, \eqref{p2:eq:LASSO} can be written in different equivalent forms. In this paper, we always refer to \eqref{p2:eq:LASSO} as the canonical form.

LASSO and group-LASSO can be numerically solved in polynomial time by off-the-shelf optimization techniques such as interior point. There is also a variety of different heuristic numerical methods to decrease its computational complexity \cite{Homotopy,LARS}. It has been observed that all of these numerical techniques run into numerical problems when LASSO is applied to a dense grid with small $\lambda$. Thus, a relatively coarse grid is applied in practice, which leads to the so-called off-grid problem as the true parameters can be relatively distant from the grid points. There has been different attempts to overcome the off-grid effect \cite{zhu2011sparsity,tang2012compressive}. To the best of our knowledge no related general study is available. 
A natural attempt to isolate the effect of discretization in an analysis is to extend LASSO to admit a continuum instead. This has been central in many recent studies such as \cite{candes2012towards,bhaskar2011atomic,tang2012compressive}, where the idea of atomic norm regularization is proposed and limited numerical approaches are discussed. In \cite{Panahi2014Gridless} an extension with a general implementation is also proposed. The numerical implementation is not a concern in the current study and as we shortly discuss, all the above extensions are theoretically equivalent. Here, we develop this unique extension again as a parametric method instead, and rename it as Continuous LASSO (CLASS) to emphasis on its parametric aspects. Then, we first relate the original LASSO with discretization to CLASS by an asymptotic analysis and next analyze CLASS.   

\subsection{Preliminaries on Asymptotic Analysis}

The analysis herein is carried out based on some asymptotic assumptions. We have already pointed out the issue of grid density. Here we clarify the assumptions under which the analysis holds. In short, our analysis admits a case with sufficiently dense grid and highly small noise variance $\sigma$. To formulate the density of a grid, one may find the following definition useful.
   
\begin{Definition}\label{p2:def:density}
A finite grid $G\subset\Theta$ is called $\delta-$dense if for any $\theta\in\Theta$ there exists a close sample $\ttheta^k\in G$ such that $|\theta-\ttheta^k|<\delta$. In other words,
\begin{equation}
\max\limits_{\theta\in\Theta}\min\limits_{\ttheta^k\in \tilde{\Theta}}|\theta-\ttheta^k|<\delta
\end{equation}
\end{Definition}

The asymptotic analysis of LASSO over dense grids, connects it to CLASS. We shortly show that the group-LASSO estimate for a $\delta-$dense grid approaches a fixed solution when $\delta$ tends to zero, i.e. when the grid is densified. The limit, CLASS, is independent of other properties of the grid, and coincides with the result of the approaches in \cite{candes2012towards,bhaskar2011atomic,tang2012compressive}. Although one may find this result intuitively trivial, the real difficulty here is in the mathematical development, which once accomplished, provides us with a strong tool to take further steps.

The main issue with the mathematical development is that LASSO provides a solution of varying, but not always desirable order. In practice, the off-grid effect leads to a remarkably overestimated order, where each true parameter is replaced by a group of nearby estimates, later referred to as a cloud. Thus, an extra care should be taken on evaluating the accuracy of the estimates.  We define a proper distance (or more formally a topology) on the space of atomic decompositions, satisfying our practical concerns. This is given in Appendix \ref{p2:appendix:topology}.
 
We also consider a high Signal-to-Noise Ratio (SNR) scenario, where the ideal decomposition is close to the noiseless ADP in \eqref{p2:eq:premodel}. For the noiseless case, the LS term in LASSO can be replaced by an equality constraint, which simplifies  \eqref{p2:eq:LASSO} to
\begin{eqnarray}\label{p2:eq:LASSOnl}
&\min\limits_{\{\ts_k(t)\}}
\|\{\ts_k(t)\}\|_{2,1}\nonumber\\
&\text{s.t}\nonumber\\
&\bx(t)=\sum\limits_{k=1}^N\as(\ttheta^k)\ts_k(t)
\end{eqnarray}
This is known as the noiseless group-LASSO. It is well-known that the solution of the noiseless LASSO is also the limit solution of \eqref{p2:eq:LASSO} when $\lambda$ approaches zero. This so-called \emph{homotopy} rule suggests that in a high-SNR scenario a small value of $\lambda$ should be utilized. We later discuss selection of $\lambda$ in more detail. 

Let us summarize the above. Considering the asymptotic high SNR analysis, our strategy is to first verify if a noiseless synthesis in \eqref{p2:eq:premodel} can be recovered perfectly by the noiseless LASSO and then analyze the estimate by Taylor expansion for a small amount of noise. Note that to overcome the discretization effect, we eventually need to instead characterize the limit estimates for infinitely dense grids, called CLASS estimates. 
We also find the following well-known result, characterizing the solution of \eqref{p2:eq:LASSO} and \eqref{p2:eq:LASSOnl}  useful in our analysis. Note that group-LASSO is convex and thus its local optimality condition by the Karush-Kuhn-Tucker (KKT) theorem guaranties global optimality. The following theorem provides the KKT condition for \eqref{p2:eq:LASSO} and \eqref{p2:eq:LASSOnl}, which characterizes the LASSO solution.

\begin{fact}\label{p2:fact:KKT}
Consider a sequence $\{\ts_k(t)\}$ 
and define
\begin{equation}
\tp_k=\sqrt{\sum\limits_t|\ts_k(t)|^2}
\end{equation}
 Then, $\{\ts_k(t)\}$ is an optimal point in \eqref{p2:eq:LASSOnl} if and only if there exists a \emph{dual verifier sequence} $\bz(1),\bz(2),\ldots,\bz(T)\in\mathbb{C}^m$ such that
\begin{equation}\label{p2:eq:KKT1}
\tp_k \neq 0\to\as^H(\ttheta^k)\bz(t)=\frac{\ts_k(t)}{\tp_k}
\end{equation} 
and
\begin{equation}
\sqrt{\sum\limits_{t=1}^T\left|\as^H(\ttheta^k)\bz(t)\right|^2}\leq 1,
\end{equation}
for $k=1,2,\ldots,N$. Moreover, $\{\ts_k(t)\}$ is a solution to \eqref{p2:eq:LASSO} if the dual verifiers also satisfy 
\begin{equation}
\lambda\bz(t)=\bx(t)-\sum\limits_{k=1}^N\as(\ttheta^k)\ts_k(t)
\end{equation}
\begin{proof}
First consider the noiseless case in \eqref{p2:eq:LASSOnl}. Take $\{\ts_k(t)\}$ as the optimal solution. Applying KKT theorem, we obtain that there exists a sequence of vectors $\bz(1),\bz(2),\ldots,\bz(T)\in\mathbb{C}^m$ such that for each $k$,
\begin{equation}\label{p2:eq:2mid}
\{\as^H(\ttheta^k)\bz(t)\}_{t=1}^T\in\partial\|\{\ts_k(t)\}\|_2
\end{equation} 
where $\partial\|\{\ts_k(t)\}\|_2$ denotes the subdifferential of the multivariable 2-norm function $\|\{\ts_k(t)\}\|_2$. Note that
\begin{equation}\label{p2:eq:1mid}
\partial\|\{\ts_k(t)\}\|_2=\left\{\begin{array}{lc}
\left\{\{\eta_k(t)\}_{t=1}^T\mid\sum\limits_{t=1}^T|\eta(t)|^2\leq 1\right\} & \tp_k=0\\
\left\{\{\eta_k(t)=\frac{\ts_k(t)}{\tp_k}\}_{t=1}^T\right\} & \tp_k\neq 0
\end{array}\right.
\end{equation}
Plugging \eqref{p2:eq:1mid} into \eqref{p2:eq:2mid} proves the theorem. The noisy case in \eqref{p2:eq:LASSO} is similarly proved by the KKT Theorem and using \eqref{p2:eq:1mid}.
\end{proof}
\end{fact}  

\subsection{CLASS Solution}

Now, we show that the solutions to LASSO \eqref{p2:eq:LASSO} and \eqref{p2:eq:LASSOnl} have unique limits when the density of the grid $G$ increases. We first introduce the limit as a solution to a parametric optimization, called CLASS, and then show convergence. The main idea is that CLASS somehow generalizes LASSO to the case of a continuum. Of course, implementation aspects of CLASS falls beyond  our concern as CLASS serves only as an analytical asymptotic tool. However, we remind once again that CLASS has also been considered by other researchers as a numerical approach and the implementation aspects of CLASS is an ungoing research.

The CLASS formalism relies on the fact that any sparse decomposition can be expressed by its parameters over the support only, since other parameters are zero and uninteresting according to the principle of sparsity. Take an arbitrary collection $\{\ts_k(t)\}$ in the search space of LASSO over a fixed grid $\tilde{\Theta}=\{\ttheta^k\}$. Remember that this more precisely corresponds to the atomic decomposition $\{\ttheta^k,\{\ts_k(t)\}\}$, where $\{\ttheta^k\}$ is always neglected as it is known. Define its support $\{\theta_1=\ttheta^{k_1},\theta_2=\ttheta^{k_2},\ldots,\theta_n=\ttheta^{k_n}\}$ and note that the reduced atomic decomposition $\{\theta_l=\ttheta^{k_l},\{\ts_{k_l}(t)\}\}$ entirely represents the original decomposition. Thus, the search space of LASSO can be equivalently represented by the space of all possible reduced decompositions, i.e. the space of all irreducible representations $\{\theta_l,\{s_l(t)\}\}$ with $\theta_l\in \tilde{\Theta}$. Let us denote this space by $\tilde{\calM}$. Then LASSO can be written as

\begin{equation}\label{p2:eq:LASSOparametric1}
\min\limits_{\tilde{\calM}}\frac{1}{2}\sum\limits_{t=1}^T\left\|\bx(t)-\sum\limits_{k=1}^n\as(\theta_k)s_k(t)\right\|_2^2
+\lambda\|\{s_k(t)\}\|_{2,1}
\end{equation}
where 
\begin{equation}\label{p2:eq:norm}
\|\{s_k(t)\}\|_{2,1}=\sum\limits_{k=1}^n\sqrt{\sum\limits_{t=1}^T|s_k(t)|^2}
\end{equation}
Now, the generalization comes with relaxing the requirement that $\theta_k$ lies on the grid $\tilde{\Theta}$. Then, $\tilde{\calM}$ is replaced by the set $\calM$ of all decompositions over $\Theta$:
\begin{equation}\label{p2:eq:LASSOparametric}
\min\limits_{\calM}\frac{1}{2}\sum\limits_{t=1}^T\left\|\bx(t)-\sum\limits_{k=1}^n\as(\theta_k)s_k(t)\right\|_2^2
+\lambda\|\{s_k(t)\}\|_{2,1}
\end{equation}
We call \eqref{p2:eq:LASSOparametric} parametric LASSO or Continuous LASSO when $\Theta$ is a continuum. While the reader may verify by simple calculations that the above is a different representation of the atomic norm denoising technique introduced in \cite{bhaskar2011atomic}, it is also simple to show that the total variation formalism in \cite{candes2012towards} always leads to the same result as CLASS. Still, it is not clear that CLASS has a solution, since there is no restriction on the order $n$ and the cost  may decrease unboundedly. An independent argument on the existence of the CLASS solution is included in Appendix \ref{p2:appendix:exist} to point out to some other useful technical facts. However, the reader may refer to \cite{bhaskar2011atomic} as well. Thus, we may consider that the solution of CLASS exists. One can similarly consider the following parametric extension of the noiseless LASSO in \eqref{p2:eq:LASSO}, which we call noiseless CLASS:
\begin{eqnarray}\label{p2:eq:LASSOnlparametric}
&\min\limits_{\calM}\|\{s_k(t)\}\|_{2,1}\nonumber\\
&\text{s.t}\nonumber\\
&\bx(t)=\sum\limits_{k=1}^n\as(\theta_k)s_k(t)
\end{eqnarray}

Now, we  state the convergence theorem, which ties CLASS to the conventional LASSO with a finite grid. We actually provide a stronger convergence property which includes all later asymptotic concerns, including the noise effect and the regularization parameter. 
\begin{theorem}
\label{p2:theorem:convergence}
Consider a regular manifold $\ba(\theta)$, arbitrary observations $\bx(t)$, perturbations $\{\bn(t)\}$, a grid $\tilde{\Theta}$ and $\lambda>0$. For any desired precision $\epsilon>0$, there exists a positive real $\delta$ such that if $\|\bn(t)\|_2\leq\delta$,  $\tilde{\Theta}$ is $\delta-$dense and $\lambda<\delta$, then any group-LASSO estimate for $\{\bx(t)+\bn(t)\}$ by $\tilde{\Theta}$ and $\lambda$ are in an $\epsilon-$neighborhood of a noiseless CLASS estimate of $\{\bx(t)\}$.
\begin{proof}
See Appendix \ref{p2:appendix:homotopyproof}.
\end{proof}
\end{theorem}
 In simple words the solution of LASSO with a dense grid, small noise and regularization parameter is in the sense discussed in Appendix \ref{p2:appendix:topology} close to the noiseless solution.
\subsection{Dual Convergence Properties}

Theorem \ref{p2:theorem:convergence} shows that the solution to the noisy group LASSO is arbitrarily close to the ideal noiseless CLASS in an asymptotic case. However, to analyze LASSO in the asymptotic case, we need to characterize these solutions. We have already done this for LASSO in Lemma \ref{p2:fact:KKT}. Here, we extend this to the CLASS solution an provide convergence properties for the dual verifier vectors $\{\bz(t)\}$. Once we provide these results we can characterize small perturbations by Taylor expansion, which is discussed in the next section.
\begin{theorem}
\label{p2:theorem:CLASSdual}
A decomposition such as $\{(\theta_k,\{s_k(t)\})\}_{k=1}^n$ is a solution to noiseless CLASS with $\{\bx(t)\}$ if and only if defining $p_k=\sqrt{\sum_t|s_k(t)|^2}$, there exists a sequence of dual verifier vectors $\{\bz(t)\}$ such that
\begin{equation}\label{p2:eq:cccond1}
\ba^H(\theta_k)\bz(t)=\frac{s_k(t)}{p_k}
\end{equation}
and 
\begin{equation}\label{p2:eq:cccond2}
\forall\theta\in\Theta\quad \sum\limits_{t=1}^T|\ba^H(\theta)\bz(t)|^2\leq 1
\end{equation}
Furthermore, for each arbitrary precision $\epsilon$ there exists $\delta>0$ such that if $\tilde{\Theta}$ is $\delta-$dense, $\lambda<\delta$ and $\|\bn(t)\|_2<\delta$ then any set of dual verifiers $\{\bz_0(t)\}$ for their corresponding group LASSO over $\{\bx(t)+\bn(t)\}$ satisfies $\|\bz(t)-\bz_0(t)\|_2<\epsilon$ for a set of dual verifiers  $\{\bz(t)\}$ corresponding to a solution of noiseless CLASS.
\begin{proof}
The proof is given in Appendix \ref{p2:appendix:convergence}.
\end{proof}
\end{theorem}

\subsection{First Order Linearization}

We finally arrive at the crucial step of calculating the approximate LASSO error in a high SNR and dense grid case. We later develop conditions under which, the true parameters are exactly identical to the solution of noiseless CLASS. For the time being, we treat the noiseless CLASS solution as the desired estimate. Thus, the error is only associated with noise, grid and regularization parameter $\lambda$. Then, Theorem \ref{p2:theorem:convergence} shows that the error is infinitesimal in the vicinity of the ideal setup, i.e. when noise and $\lambda$ are small and the grid is dense. This allows for the application of a Taylor expansion. However, due to the unfamiliar role of the grid and the unspecified order of the estimates, a careful study is necessary. Let us start from the result of Theorem \ref{p2:theorem:convergence}. Take a solution $A=\{(\theta_k,\{s_k(t)\})\}$ corresponding to a $\delta-$dense grid $G$, $\delta-$small noise terms $\bn(t)$  and $\lambda<\delta$. Suppose that $\delta$ is small such that Theorem \ref{p2:theorem:convergence} guarantees that $A$ is in an $\epsilon-$neighborhood of a noiseless CLASS solution $A_0=\{(\theta_{l,0},\{s_{l,0}(t)\})\}$. The definition of neighborhood allows that some indexes of $A$, associated to an infinitesimal amplitude, lie outside the $\epsilon$-neighborhood of the elements of $A_0$. We call them \emph{false alarm}. More formally, an index $\theta_k$ is a false alarm if $|\theta_k-\theta_{l,0}|>\epsilon$ holds for all $l$. Clearly, for such an index $|s_k(t)|<\epsilon$ also holds. Note that the definition of false alarm depends on the neighborhood size $\epsilon$. Should there be a risk of confusion, we may refer to the term $\epsilon-$false alarms for clarity. The other estimates, also called "detections" (or $\epsilon-$detections) can be assigned uniquely to a close noiseless estimate in a sufficiently small neighborhood. However, the definition of neighborhood also allows for multiple detections assigned to the same index. We call this the dispersion effect, which might be related, for example, to discretization. Finally, the detections related to the same index are somehow subject to an overall estimation error (shift). Our analysis will characterize the above three asymptotic elements of estimation; false alarm, dispersion and the overall estimation error.

To formulate the asymptotic behavior of LASSO in the above sense, we first need to review some basic definitions. 
Consider again the above solution $A$ in a sufficiently small $\epsilon-$neighborhood of the noiseless solution $A_0$ such that each index $\theta_k$ in $A$ is either a false alarm or is uniquely located in an $\epsilon-$neighborhood of an index $\theta_{l,0}$. We refer to all elements $\theta_k$ in the neighborhood of a specific element $\theta_{l,0}$ as its corresponding \emph{cloud}. We basically show that each cloud may consist of at most 2 elements of zero or first order. To elaborate on this, consider the third largest element in each cloud and denote the maximum amplitude of these elements by $\delta_3$. Then, we show that $\delta_3$ vanishes up to first order with respect to $\delta$. Finally, we define the "overall" effect of each cloud by the following parameters:
\begin{subequations}
\begin{eqnarray}
&\sigma_l(t)=\sum\limits_{k\mid|\theta_k-\theta_{0,l}|<\epsilon}s_k(t)-s_{l,0}(t)\\
&\pi_l=\frac{1}{p_{l,0}}\sum\limits_{k\mid|\theta_k-\theta_{0,l}|<\epsilon}p_k(\theta_k-\theta_{l,0})
\end{eqnarray} 
\end{subequations}
where  $p_k=\sqrt{\sum\limits_t|s_{k}(t)|^2}$ and $p_{l,0}=\sqrt{\sum\limits_t|s_{l,0}(t)|^2}$.
In fact, it is simple by Taylor expansion to see that the first order properties of any estimator is well expressed by the above parameters, where $\sigma_k(t)$ is complex-valued and $\pi_k(t)$ is real-valued. Note that, in general the characteristics of $\sigma$ and $\pi$ do not completely reveal the properties of individual indexes and amplitudes in each cloud
, which after all, depend on the circumstances (e.g. discretization) under which the cloud is produced.
Now, define
\begin{equation}\label{p2:eq:gfunction}
g=\frac{1}{2}\sum\limits_t\left\|\bn(t)-\sum\limits_l\left(\ba_l\sigma_l(t)+\bd_ls_{l,0}(t)\pi_l\right)-\sum\limits_p\ba(\bar{\theta}_p)\bar{s}_p(t)\right\|_2^2
+\lambda\sum\limits_p\sqrt{\sum\limits_t|\bar{s}_p(t)|^2}+\lambda\sum\limits_{l,t}\Re(\gamma^*_l(t)\sigma_l(t))
\end{equation} 
where
\begin{eqnarray}
&\ba_l=\ba(\theta_{l,0})\nonumber\\
&\bd_l=\frac{\text{d}\ba}{\text{d}\theta}(\theta_{l,0})\nonumber\\
&\gamma_l(t)=\frac{s_{l,0}(t)}{\sqrt{\sum\limits_t|s_{l,0}(t)|^2}}
\end{eqnarray}
This is a function of $\{\pi_l,\{\sigma_l(t)\}\}$ and an arbitrary decomposition $\bar{A}=\{\bar{\theta}_p,\{\bar{s}_p(t)\}\}$. The following theorem identifies the first order perturbation of the solution in terms of the above definitions.

\begin{theorem}\label{p2:theorem:pert} 
Consider LASSO with a $\delta-$dense grid, and $\lambda<\delta$ over observations with small perturbation $\|\bn(t)\|<\delta$ such that any solution $A$ lies in a small $\epsilon-$neighborhood of a noiseless solution $A_0$.

{\bf a)} Minimizing $g$ in \eqref{p2:eq:gfunction} gives the first-order perturbation of the noisy solution:  Consider the optimization
\begin{equation}\label{p2:eq:perturbation}
\min\limits_{\{\pi_l\in\bbR,\{\sigma_l(t)\bbC\}\},\{\bar{\theta}_p,\{\bar{s}_p(t)\}\}}g
\end{equation}
 There exists a minimum point $\bar{\pi}_l,\bar{\sigma}_l(t)$ and $\bar{A}$ such that up to first order, $\pi_l,\sigma_l(t)$ and false alarms are identical to $\pi_l,\sigma_l(t)$ and $\bar{A}$.

{\bf b)} There exists a solution of LASSO for which the maximum false amplitude $\delta_3$ vanishes up to first order, i.e. $\delta_3=o(\delta)$.
\begin{proof}
See Appendix \ref{p2:appendix:linearization} for proof and more details.
\end{proof}
\end{theorem}  

Theorem \ref{p2:theorem:pert} may be regarded as the central contribution of this work. Once this is established, characterizing the high SNR properties of LASSO boils down to analyzing the minimizers of the linearized criterion $g$. The next section provides such an analysis, where we use the linearization result to give a statistical analysis of LASSO estimates in presence of a white Gaussian noise.

\section{Statistical Results}
\label{p2:sec:stat}
In the previous section, we developed results characterizing the LASSO estimates in an asymptotic case. In this section, we connect those results to practice. We shortly address the statistical effect of noise and grid on the estimation procedure. We also deal with a more fundamental question of consistency. Recall that the linearization results characterize the deviation from the noiseless solution, but we have not yet discussed the own properties of the noiseless solution. Many previous studies have considered this and what we correspondingly state in the sequel is more or less a restatement of the results in \cite{candes2012towards,tang2012compressive} for special cases, which is derived more systematically as a part of a general framework resulting from Theorem \ref{p2:theorem:pert}. In fact, Theorem \ref{p2:theorem:pert} is central in the entire discussion of the current section, which readily characterizes the first order deviation from a noiseless solution. It only remains to investigate the statistical properties of the deviation in a given scenario. Hence, it seems rational to spend a bit of effort first to learn more about the consequences of Theorem \ref{p2:theorem:pert}.

Let us start by some simplifying definitions. Consider a noiseless solution $A=\{\theta_k,\{s_k(t)\}\}$ and its corresponding parameters $\ba_k=\ba(\theta_k)$, $\bd_k=d\ba/d\theta(\theta_k)$ and $\gamma_k(t)=s_k(t)/p_k$, where $p_k=\sqrt{\sum_t|s_k(t)|^2}$. Define $\bA=[\ba_1\ \ba_2\ \ldots \ba_n]$ and $\bD=[\bd_1\ \bd_2\ \ldots \bd_n]$ as well as $\bA^\dagger=(\bA^H\bA)^{-1}\bA^H$ and $\bP=\bI-\bA\bA^\dagger$, where $\bI$ denotes the unit matrix. Finally, define
\begin{eqnarray}
&\xi_{l,k}=\sum\limits_{t=1}^Ts_{l,0}^*(t)s_{k,0}(t)\nonumber\\
&\bR=\Re\left[(\bD^H\bP\bD)\odot\Xi\right]
\end{eqnarray}
where $\odot$ denotes elementwise product and $\Xi$ is the matrix of the elements $\xi_{l,k}$. Denote by $\bxi_k$ the $k^{\text{th}}$ column of $\Xi$. 

Now let us try to solve \eqref{p2:eq:perturbation}. Note that fixing the false alarm $\bar{A}$, the optimization over the $\pi$ and $\sigma$ parameters is quadratic of the following form
\begin{equation}
\min\limits_{\{\pi_l(t),\sigma_l(t)\}}
\frac{1}{2}\sum\limits_t\left\|\bnu(t)-\sum\limits_l(\ba_l\sigma_l(t)+\bd_ls_{l,0}(t)\pi_l))\right\|_2^2
+\lambda\sum\limits_{l,t}\Re(\gamma^*_l(t)\sigma_l(t))
\end{equation}
where
\begin{equation}
\bnu(t)=\bn(t)-\sum\limits_p\ba(\bar{\theta}_p)\bar{s}_p(t)
\end{equation}
and the constant terms are neglected. The solution to this can easily be found by differentiation as
\begin{eqnarray}\label{p2:eq:pisigma}
&\bsigma(t)=\bA^\dagger(\bnu(t)-\sum\limits_l\bd_ls_{l,0}(t)\pi_l))-\lambda(\bA^H\bA)^{-1}\bgamma(t)\\
&\bpi=\bR^{-1}(\bomega+\lambda\bdelta)
\end{eqnarray}
where $\bsigma(t),\bpi,\bomega$ and $\bdelta$ denote the vectors with $\sigma_k(t),\pi_k,\omega_k$ and $\delta_k$ as elements respectively such that
\begin{equation}
\omega_k=\Re(\bzeta_k^H\bP\bd_k)\quad\delta_k=\Re(\bxi_k^T\bA^\dagger\bd_k)
\end{equation}
where
\begin{equation}
\bzeta_k=\sum\limits_{t=1}^Ts_{k,0}^*(t)\bnu(t)
\end{equation}
On the other hand, fixing $\{\pi_k,\{\sigma_k(t)\}\}$, the optimization over  false alarm is a LASSO problem. We have found it both difficult and practically uninteresting to fully analyze the properties of the false alarm solution as a random atomic decomposition, or more restrictively, a random finite set. Instead, we only study the occurrence of false alarm, which is to identify when $\bar{A}$ is nonempty in \eqref{p2:eq:perturbation}. Note that when $\bar{A}$ is empty $\bnu(t)=\bn(t)$ and according to Theorem \ref{p2:theorem:CLASSdual} the following relation equivalently holds for any $\theta$.
\begin{equation}\label{p2:eq:FAcond}
\sum\limits_t\left|\ba^H(\theta)\left(\bn(t)-\sum\limits_l(\ba_l\sigma_l(t)+\bd_ls_{l,0}(t)\pi_l))\right)\right|^2\leq\lambda^2
\end{equation}
where $\sigma_k(t)$ and $\pi_k$ are given by \eqref{p2:eq:pisigma}. We define the probability of false alarm (PFA) as $\text{PFA}=\text{Pr}(\bar{A}\neq \emptyset)$. 

\subsection{Ideal Consistency}
Based on the above, we now provide a sufficient condition for a true decomposition to be exactly retrieved by noiseless CLASS. Note that Theorem \ref{p2:theorem:CLASSdual} gives a necessary and sufficient condition for our purpose. However, it is not straightforward to verify it by introducing dual verifiers. What we do in the sequel is to propose a certain choice of dual verifiers based on \eqref{p2:eq:FAcond} which is easier to verify and still identifies a large class of consistent cases. 

Considering \eqref{p2:eq:FAcond}, it is easy to recognize that some individual terms depend on noise and vanish in the noiseless case. This concerns a case where the noise data are processed by noisy LASSO, a dense grid and small $\lambda$. Then,
\begin{eqnarray}\label{p2:eq:pisigmanl}
&\bsigma(t)=-\lambda\bA^\dagger(\sum\limits_l\bd_ls_{l,0}(t)\pi_{0,l}))-\lambda(\bA^H\bA)^{-1}\bgamma(t)=\lambda\bsigma_0(t)\\
&\bpi_0=\bR^{-1}\bdelta
\end{eqnarray}
where $\bpi_0$ is a vector consisting of the elements $\pi_{0,l}$ and \eqref{p2:eq:FAcond} can also be characterized by
\begin{equation}\label{p2:eq:FAcondnl}
\lambda^2\sum\limits_t\left|\ba^H(\theta)\left(\sum\limits_l\left(
\bP\bd_l\gamma_l(t)\pi_{0,l}-\bA(\bA^H\bA)^{-1}\bgamma(t)\right)
\right)\right|^2\leq\lambda^2
\end{equation}
As \eqref{p2:eq:FAcond} identifies existence of false alarm, \eqref{p2:eq:FAcondnl} naturally identifies a case where application of noisy LASSO results in no false alarm. We call this case purely consistent. However, \eqref{p2:eq:FAcondnl} implies pure consistency only when consistency is priorly established.  Fortunately, it can also be seen that \eqref{p2:eq:FAcondnl} automatically implies consistency as the vector 
\begin{equation}\label{p2:eq:dualpure}
\bz(t)=\sum\limits_k\sigma_{k,0}(t)\ba_k+\sum\limits_k\pi_{k,0}s_{k,0}(t)\bd_k
\end{equation}
would then satisfy Theorem \ref{p2:theorem:CLASSdual} by direct calculation. The following theorem summarizes and completes the above discussion.
\begin{theorem}\label{p2:theorem:consistecy}
{\bf a}) A decomposition $A=\{\theta_k,\{s_k(t)\}\}$ is consistent, i.e. LASSO estimates for its corresponding observation by a sufficiently small noise is arbitrarily close to $A$ if \eqref{p2:eq:FAcond} holds, in which case it is also purely consistent.

{\bf b}) Any consistent decomposition is a subset of a purely consistent decomposition.
\begin{proof}
See Appendix \ref{p2:appendix:consistecy}.
\end{proof}

\end{theorem}

\subsection{Statistical Properties of Perturbations}

Let us assume that a consistent true decomposition is observed by the model in \eqref{p2:eq:model} and the noise perturbation is so small and the grid is so dense that Theorem \ref{p2:theorem:pert} characterizes the estimation error. Thus, we may analyze the statistical properties of the solution of \eqref{p2:eq:perturbation} to understand the statistical behavior of the LASSO solution.  It may be readily seen that the overall error properties $\pi,\sigma$ as well as PFA are linked to false alarm, which subsequently depends on the choice of $\lambda$. On the other hand, the method of selecting $\lambda$ is not inherent in the machinery of LASSO. Thus, we examine the previous results in terms of an arbitrary $\lambda$ in some example cases. As already stated, the results are given in terms of the $\pi,\sigma$ parameters and PFA.

We will discuss in the following two different cases of interest. In the first case, the true order is known and $\lambda$ is adapted to provide an estimate of correct order. In the second one, the order is unknown and $\lambda$ is fixed to meet a certain PFA criterion. In either case, \eqref{p2:eq:FAcond} is useful as it  characterizes when no false alarm occurs.  

\subsubsection{Known Order and Adaptive Regularization}

When the number of parameters is known, $\lambda$ may be selected based on the given data set to provide a correct number of estimates. In this case no false alarm is observed and thus $\lambda$ satisfies \eqref{p2:eq:FAcond}. To investigate the best performance, we select smallest such value of $\lambda$ and denote it by $\lambda_b$. In this case, $\lambda$ becomes a function of the noise realization. Hence, it is a random variable. Remember that now $\bnu(t)=\bn(t)$. Thus, the expressions for $\pi$ and $\sigma$ and their corresponding statistics can be easily calculated. The following theorem summarizes the final expressions.

\begin{theorem}
\label{p2:theorem:statistics}
{\bf a)} The $\lambda_b$ may be calculated by
\begin{equation}\label{p2:eq:lambdab}
\lambda_b=\max\limits_{\theta}\Lambda(\theta)
\end{equation}
where $\Lambda(\theta)$ is the unique positive solution of the following equation for $\lambda$.
\begin{equation}\label{p2:eq:Lambda}
\sum\limits_t\left|\ba^H(\theta)\left(\bn(t)-\sum\limits_l(\ba_l\sigma_l(t)+\bd_ls_{l,0}(t)\pi_l))\right)\right|^2=\lambda^2
\end{equation}
and $\pi_l$ and $\sigma_l$ are given in \eqref{p2:eq:pisigma} applying $\bnu=\bn$.

{\bf b)} When the regularization parameter is selected as $\lambda_b$, the estimates have the following first-order statistical properties:
\begin{eqnarray}\label{p2:eq:stat_adaptive}
&\calE(\bpi)=\calE(\lambda_b)\bR^{-1}\bdelta\\
&\calE(\bsigma)=\calE(\lambda_b)(\bA^H\bA)^{-1}\bgamma(t)\\
&\text{Cov}(\bpi)=\bR^{-1}\Re\left[(\bD^H\bP\bC\bP\bD)\odot\Xi\right]\bR^{-1}+
\text{Var}(\lambda_b)\bR^{-1}\bdelta\bdelta^T\bR^{-1}\label{p2:eq:cov}
\end{eqnarray}
where $\bC$ denotes the covariance of noise $\bn(t)$.
\begin{proof}
The proof is given in Appendix \ref{p2:appendix:statistics}.
\end{proof}
\end{theorem}
The expression in \eqref{p2:eq:cov} has an interesting interpretation. The first term is recognized as the error covariance of the ML $\theta$ estimates \cite{stoica1989music}. The second term, proportional to the regularization parameter, is the additional contribution due to the regularization. Note that in absence of dispersion, i.e. when each parameter corresponds to single estimate, the $\pi$ parameters are equivalent to $\theta$ and thus the current analysis shows that ML is a special case of LASSO, where no dispersion and no regularization exists. We remind that in presence of dispersion, only $\pi$ parameters can be calculated by a first order approximation.        

\subsubsection{Unknown Order and fixed $\lambda$}

When the order is unknown, $\lambda$ may be fixed to set a balance between PFA and error parameters in the absence of false alarm. Although a data driven $\lambda$ is still a valid choice, it remains out of scope of the current analysis. When false alarm occurs, there is no agreed definition of the performance. Thus, we consider the average error in $\bpi,\bsigma$ only in absence of false alarm, which can be  mathematically written as
\begin{equation}\label{p2:eq:stat_fixed}
MSE_f=\calE(\bpi\bpi^T\mid NFA)
\end{equation}
where $NFA$ denotes the event that no false alarm occurs. Together with $PFA$, the above constitutes the performance measure. Unfortunately, we have not been able to provide analytical expressions for this case, since assuming $NFA$ changes the posterior distribution of $\bn(t)$ in a non-tractable way. In the next chapter we show numerical calculations based on a Monte Carlo method for this case. 

\section{Numerical Results}

We have previously formulated a parametric approach to analyze LASSO and provided the details for a high-SNR scenario. In this section, we examine our previous derivations in the case of ADP applied to DOA estimation. The numerical results can be categorized into two groups. In the first, the theoretical results are calculated by Monte Carlo techniques, reminding that some expectations could not be analytically calculated in the previous derivation. The second group compares the theoretic performance to that of some alternative methods. We consider the CLASS (atomic-norm denoising) implementation in \cite{bhaskar2011atomic, tang2012compressive}, only considering the frequency estimation problem (ULA in our case) with uniform samples. 

\subsection{Evaluation of Theoretical Performance}

\begin{figure}[t]
\centering
\includegraphics[width=15cm]{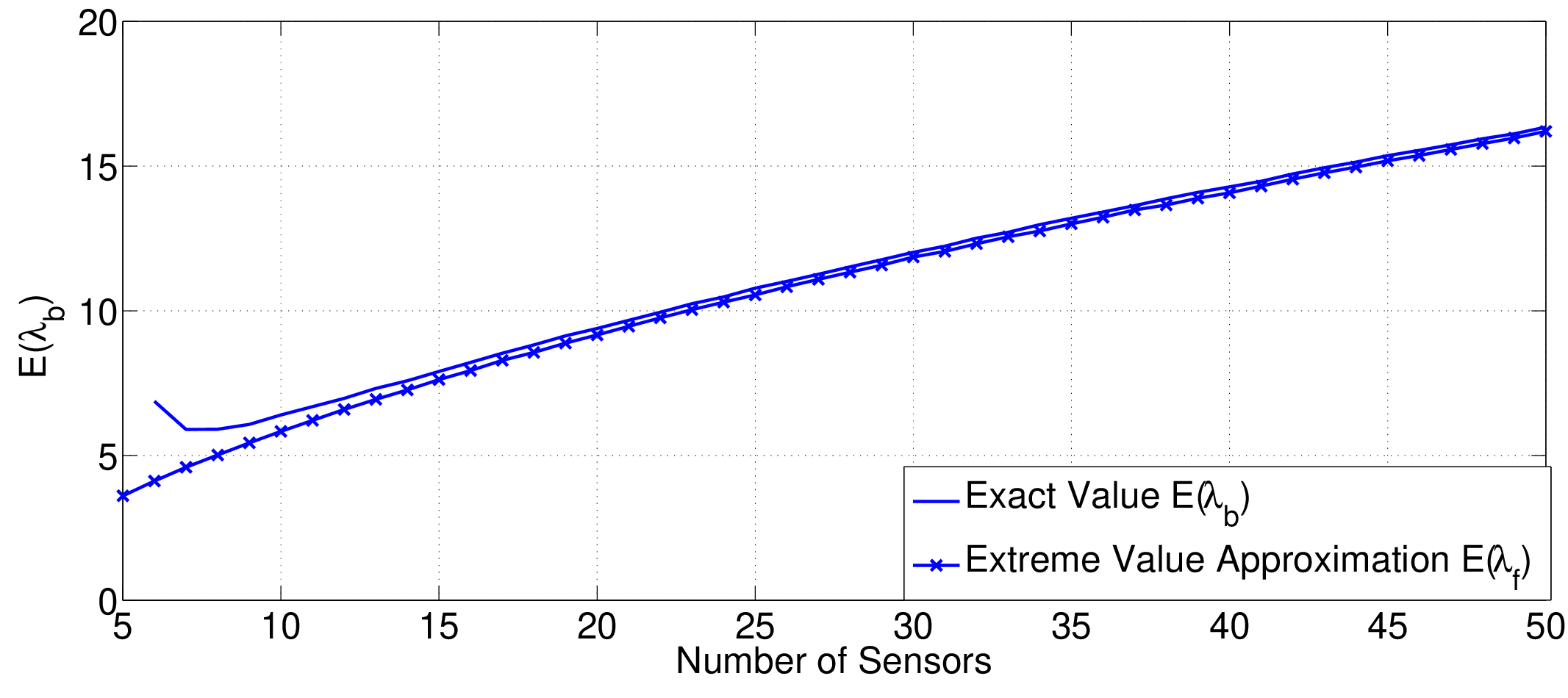}
\caption{Mean value of $\lambda_b$ compared to $\lambda_f$ for different number of sensors.} 
\label{p2:fig:lambda_E}
\end{figure}

\begin{figure}[t]
\centering
\includegraphics[width=15cm]{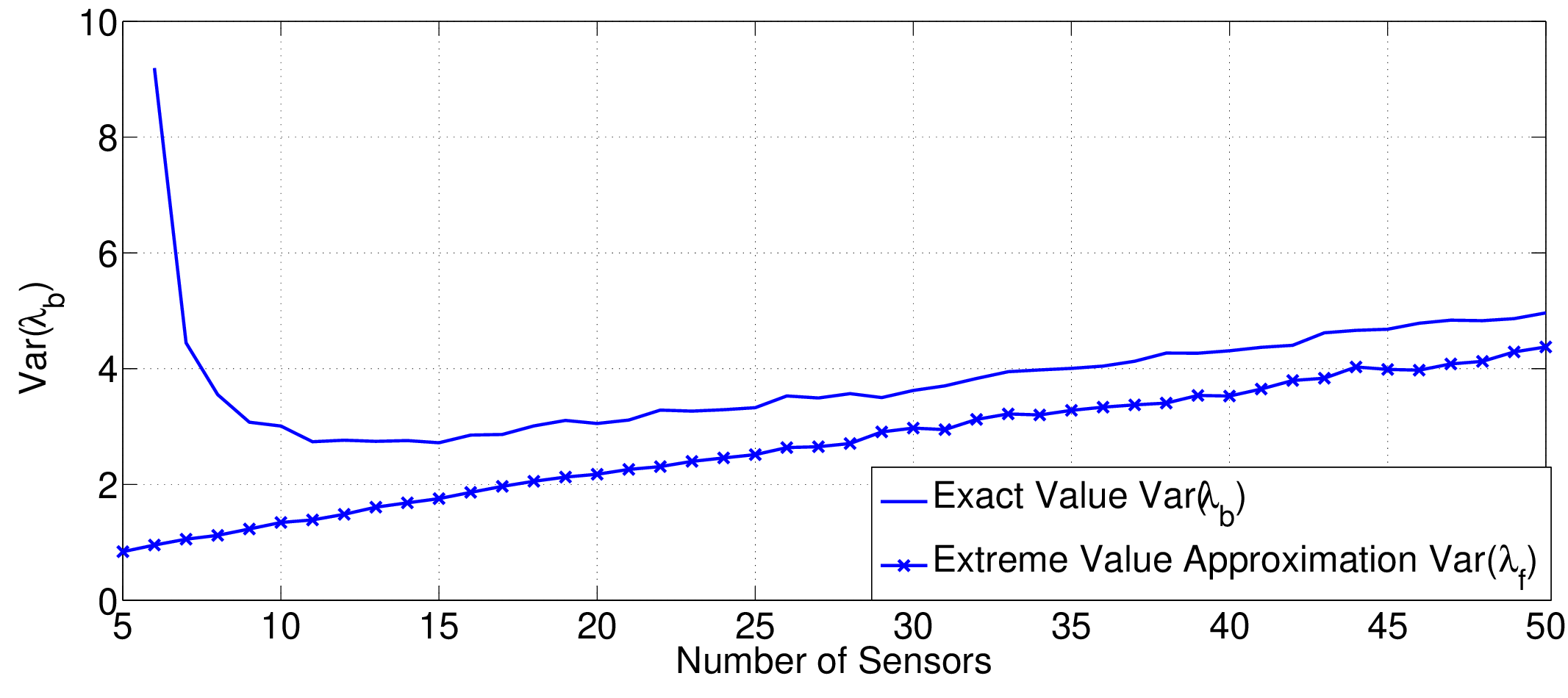}
\caption{The variance of $\lambda_b$ compared to $\lambda_f$ for different number of sensors.}
\label{p2:fig:lambda_Var}
\end{figure}

Equations \eqref{p2:eq:stat_adaptive} and the definition of $MSE_f$ and false alarm in \eqref{p2:eq:stat_fixed} and \eqref{p2:eq:FAcond}, respectively constitutes the analysis. However, evaluating them in practice needs a complicated numerical procedure. In particular, we are interested in calculating the first and second order statistics of $\lambda_b$ as well as $MSE_f$ and $PFA$ by a Monte Carlo method, which provides the results in Figures \ref{p2:fig:lambda_E} and \ref{p2:fig:lambda_Var}.

Taking a closer look at the definition of $\lambda_b$ in \eqref{p2:eq:lambdab}, one may suspect that under certain practical assumptions, many terms in \eqref{p2:eq:lambdab} can be neglected such that $\lambda_b$ can be approximated by $\lambda_f$ given by
\begin{equation}
\lambda_f=\sqrt{\max\limits_{\theta\in\Theta}\sum\limits_t|\ba^H(\theta)\bn(t)|^2}
\end{equation}
The statistics of $\lambda_f$ is widely considered in the design of Constant-False-Alarm-Rate (CFAR) estimators. Note that unlike $\lambda_b$, $\lambda_f$ is independent of the true decomposition, while still depending on the noise realization. The statistics of $\lambda_f$ can also be analytically expressed in some asymptotic cases.
Figure \ref{p2:fig:lambda_E} shows the evaluated expected value for different dimensions of observation $m$, where LASSO is applied to data from a ULA (Fourier) manifold explained in Section 2. The true DOAs are fixed at electrical angles $[0 \ 2.5\pi/m]$ with corresponding amplitudes $[1\ 1]$. The results are taken over $10000$ trials.  
Figure \ref{p2:fig:lambda_Var} shows the variance with a similar setup. As seen, $\lambda_f$ may be considered in practice as a good approximate value especially for a high number of sensors, where the relative error decreases. 

For the case of fixed $\lambda$ we calculated $MSE_f$ and $PFA$  by another MC experiment. We compared two different choices of DOA separation, namely $2.5\pi/m$ and $2.7\pi/m$, both with unit coherent amplitudes. A single snapshot was considered and the SNR and $m$ were set to $10$dB and $10$ respectively.  
\begin{figure}[t]
\centering
\includegraphics[width=15cm]{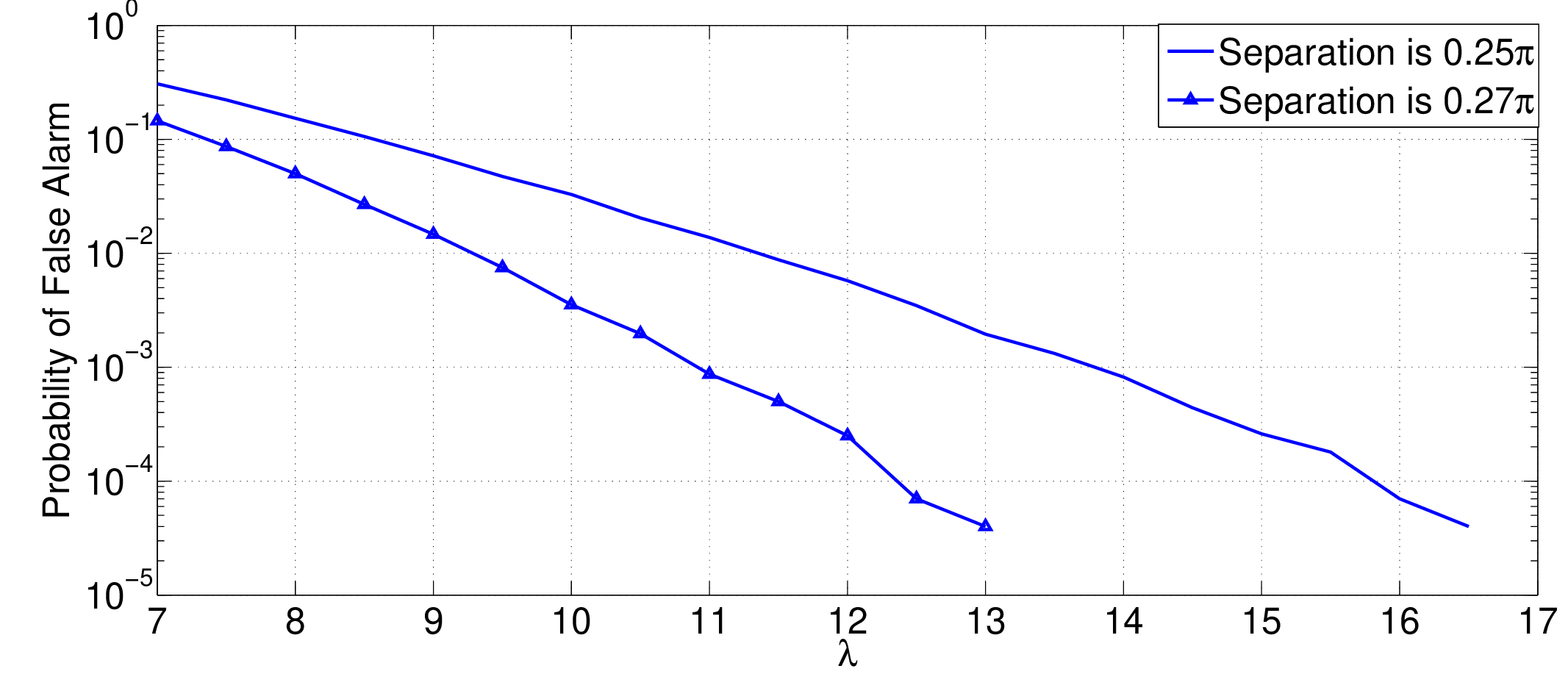}
\caption{The PFA for different values of $\lambda$ compared for different DOA separations.}
\label{p2:fig:PFA_lambda}
\end{figure}
Figure \ref{p2:fig:PFA_lambda} shows the resulting average PFA for different values of $\lambda$ over $10^5$ realizations. As seen, better separated sources need smaller value of $\lambda$ to achieve a required PFA. 
\begin{figure}[t]
\centering
\includegraphics[width=15cm]{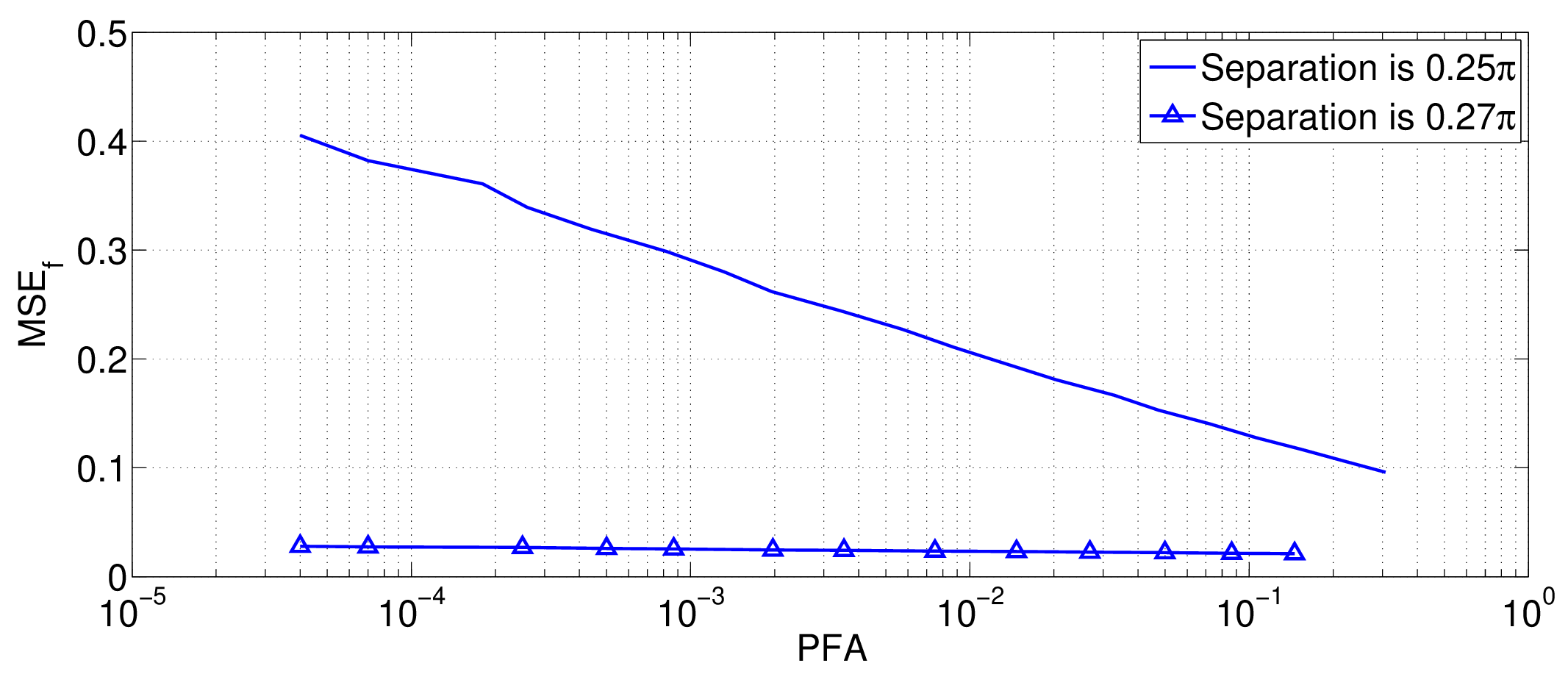}
\caption{The PFA versus MSE for different values of $\lambda$ compared for different DOA separations.}
\label{p2:fig:MSE_PFA_2.5p}
\end{figure}

Then, Figure \ref{p2:fig:MSE_PFA_2.5p} shows the trade-off between $MSE_f$ and $PFA$ in the above scenarios. 
As seen, the error dramatically decreases by increasing the separation. Reaching to the separation of $3\pi/m$, the error practically approaches the Cramer-Rao bound in the desirable range of $PFA$. The same trend is observed when the number of sensors increases from $10$ to $20$.

\subsection{Comparison with Other Methods}

\begin{figure}
\centering
\psfrag{NLLS}{ML}
\includegraphics[height=2in, width=3.5in]{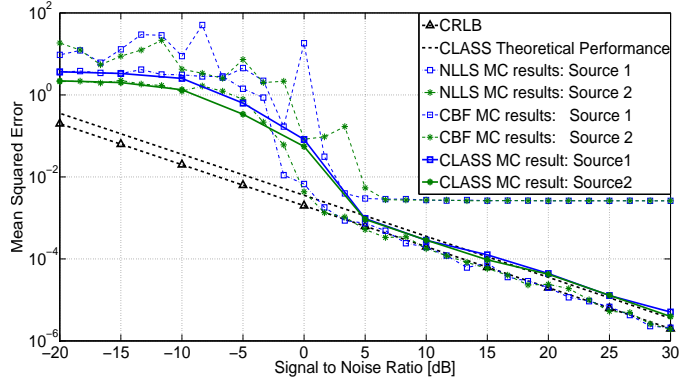}
\caption{The statistical MSE for different methods versus input SNRs. The estimation is based on one snapshot measurement of two sources separated by $\Delta\theta=\frac{4\pi}{m}$, and waveform values $s_1=s_2=1$.}
\label{p2:fig:methods_SNR}
\end{figure}

\begin{figure}[t]
\centering
\includegraphics[height=2in, width=3.5in]{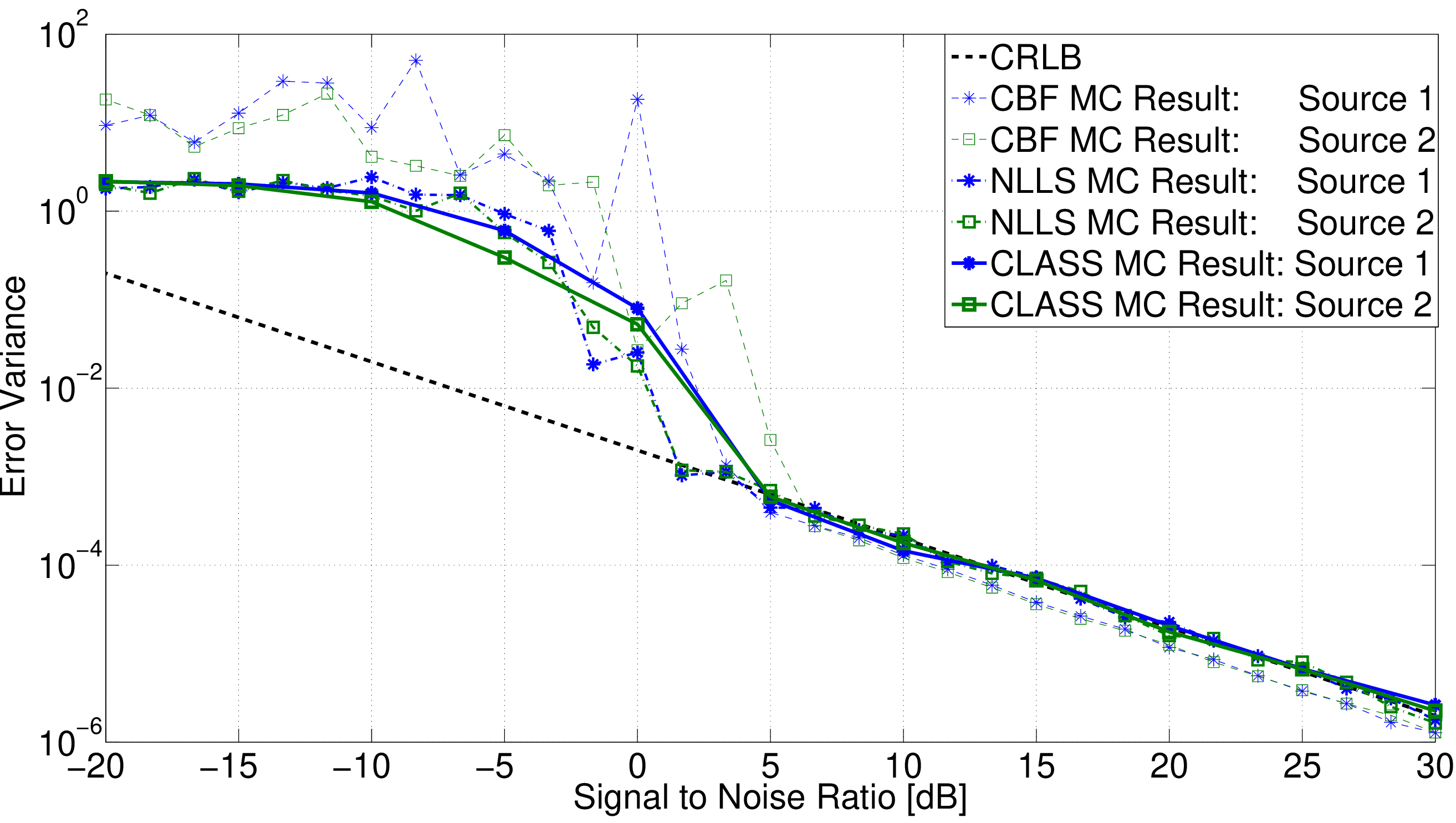}
\caption{The statistical variance for different methods in different input SNRs. The estimation is based on one snapshot measurement of two sources separated by $\Delta\theta=\frac{4\pi}{m}$, and waveform values $s_1=s_2=1$. }
\label{p2:fig:methods_variance}
\end{figure}

%
%
We finally compared the LASSO performance to that of ML (see (2))  with exhaustive search \cite{ottersten1993exact} and Conventional BeamForming (CBF) \cite{Viberg_2decades}. Figures \ref{p2:fig:methods_SNR} and \ref{p2:fig:methods_variance} compare the estimate Mean Squared Errors and variances of three different estimators; CLASS, ML and CBF, respectively. The setup is similar to the one in Figures 2 and 3, while the number of sensors $m$ is fixed to 15. The results are the average of the outcomes of 100 trials at each noise level. We see that while the asymptotic variances of CLASS and ML methods coincide, the CLASS estimator has a higher asymptotic MSE. We conclude that CLASS modifies the solution of ML mostly by adding a bias term in the very high SNR regime. However, as SNR decreases, the MSE of CLASS reaches the one for the ML estimator in the SNR regime between -2 and 5 dBs. There is a significant (almost 3 dB) difference between threshold edge of LASSO and ML. Note that ML with exhaustive search is not practical and the difference might be less with a more realistic implementation.

\section{Concluding Remarks}

This work was devoted to analysis of LASSO as a non-linear estimator of a parametric signal. The main idea here was to parametrize the support, which brought a parametric interpretation of LASSO. To meet the continuous estimation requirements, the parametric LASSO was modified to CLASS. This is similar in nature to the approach in \cite{bhaskar2011atomic}. The parametric CLASS estimates were then analyzed by linearization in a high-SNR case and related to the original estimates by LASSO. The numerical implementation of CLASS was out of the scope of the current work. However, \cite{bhaskar2011atomic} also provides a CLASS implementation for a specific case, which we employed for numerical validation.

The above approach enabled to analyze LASSO more deeply from a deterministic point of view, which is of a great interest in applications, where LASSO is utilized to estimate parameters, such as radar localization. Although, important properties of LASSO, especially the role of the RP, presented in a limited number of scenarios, the current work provides a framework for further investigations. The MSE calculations also provide a new insight to the role of the RP. With our approach, we were able to calculate MSE and the false alarm rate, which commonly characterize an estimator of varying order in the high-SNR case. The process of false alarms were more generally characterized, but we left a more accurate investigation for a future work.

The current theoretical and numerical results suggest that LASSO provides a good trade-off between error and PFA, under some considerations about resolution. This is verified for a fixed-RP scenario. However, we suspect that employing a thresholding scheme would reduce PFA more, thus further improving the properties of LASSO. However, the numerical implementation may be crucial for the performance, and should therefore be the subject of a future study.  

%
%
%
%
 
\appendices

\section{LASSO Topology on ADP space}
\label{p2:appendix:topology}
This part includes the definition of distance between atomic decompositions. Despite its complex technical definition it implies a natural concept, which easily follows from the analysis of LASSO.  

\begin{Definition}\label{p2:def:distance}
(LASSO-topology)

{\bf a)} Consider an irreducible decomposition $A=\{\{s_k(t)\},\theta_k\}_{k=1}^n$ and another arbitrary decomposition  $\bar{A}=\{\{\bar{s}_k(t)\},\bar{\theta}_k\}_{k=1}^{\bar{n}}$. 
Let $I_k=(\theta_k-\epsilon\ \theta_k+\epsilon)$ be the $\epsilon-$ball at $\theta_k$.
Then, $\bar{A}$ 
is said to be in $\epsilon-$neighborhood of $A$ if
\begin{enumerate}
\item The $\epsilon-$balls $I_k$ cover all indexes of $\bar{A}$, i.e. $\{\bar{\theta}_k\}\subset\bigcup\limits_{l=1}^nI_l$
\item For each interval $I_k$ at each time index $t$, we have that
\begin{equation}
\left|s_l(t)-\sum\limits_{k\mid\bar{\theta}_k\in I_l}\bar{s}_k(t)\right|<\epsilon
\end{equation}
\item For each $\bar{k}$ and $k$ the relation $\bar{\theta}_{\bar{k}}\in I_k$ implies that there exists $\alpha_{k,\bar{k}}>0$ such that
\begin{equation}
\forall t\ |\alpha_{k,\bar{k}}s_k(t)-\bar{s}_{\bar{k}}(t)|<\epsilon
\end{equation}
\end{enumerate}

{\bf b)} Two arbitrary decompositions $B$ and $\bar{B}$ are called $\epsilon-$similar and shown by $A\sim_\epsilon\bar{A}$ if there exists an irreducible decomposition $A$ such that both $B$ and $\bar{B}$ are in $\epsilon$-neighborhood of $A$. 

\end{Definition}
\begin{figure}
\centering
\psfrag{I1}{$I_1$}
\psfrag{I2}{$I_2$}
\psfrag{I3}{$I_3$}
\begin{subfigure}[b]{0.4\textwidth}
\includegraphics[width=7cm]{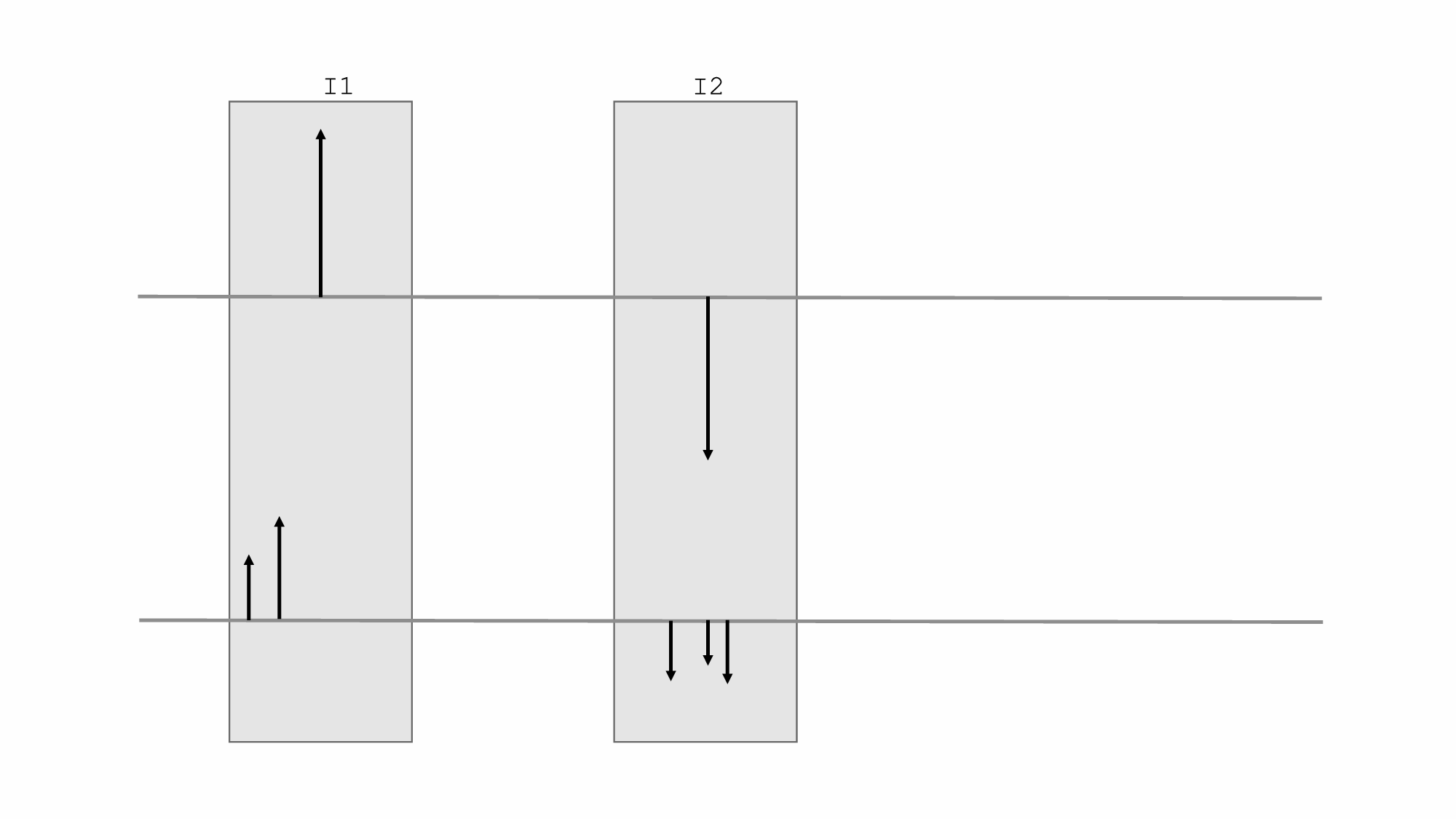}
\caption{}
\end{subfigure}
\begin{subfigure}[b]{0.4\textwidth}
\includegraphics[width=7cm]{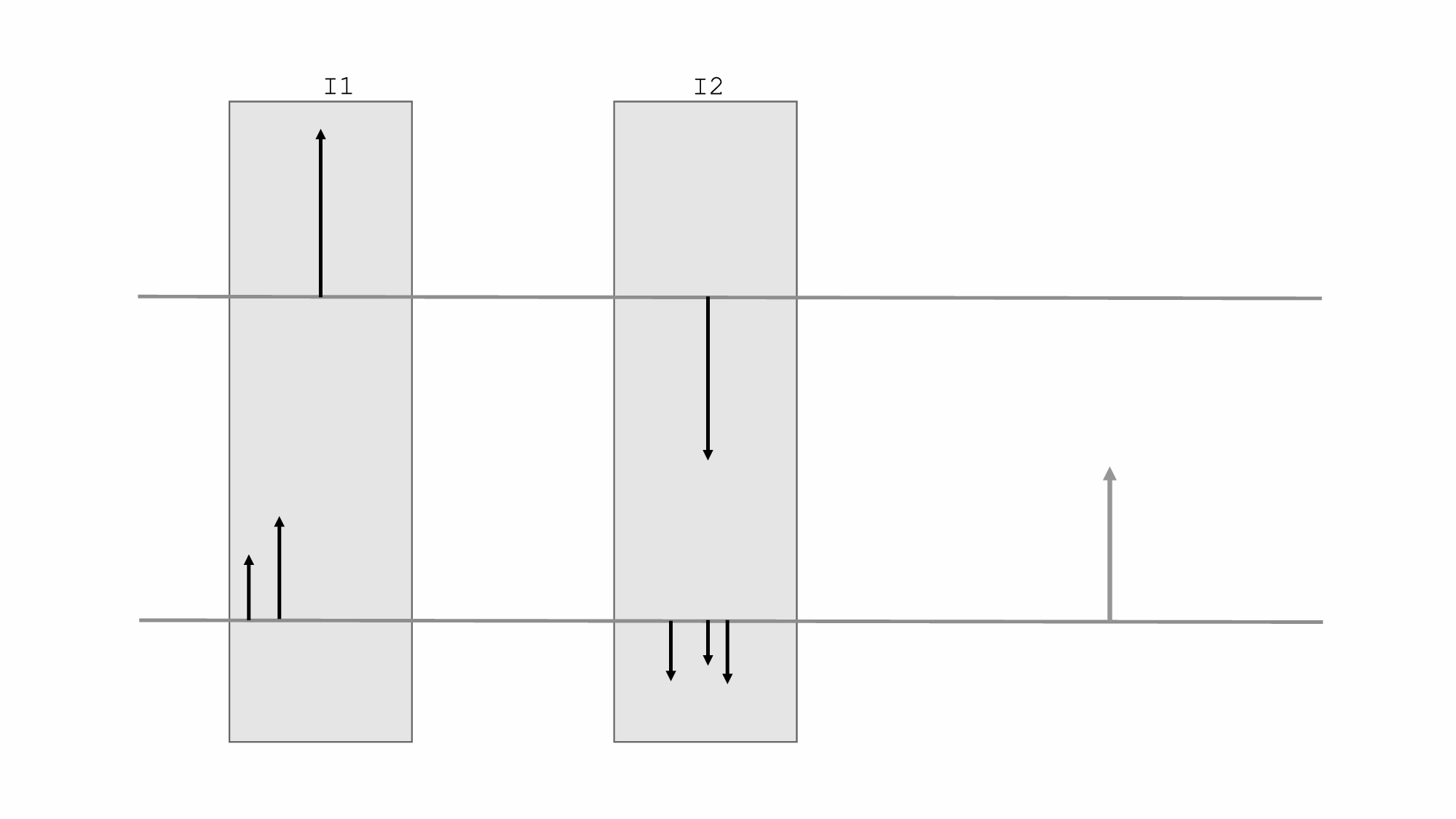}
\caption{}
\end{subfigure}
\begin{subfigure}[b]{0.4\textwidth}
\includegraphics[width=7cm]{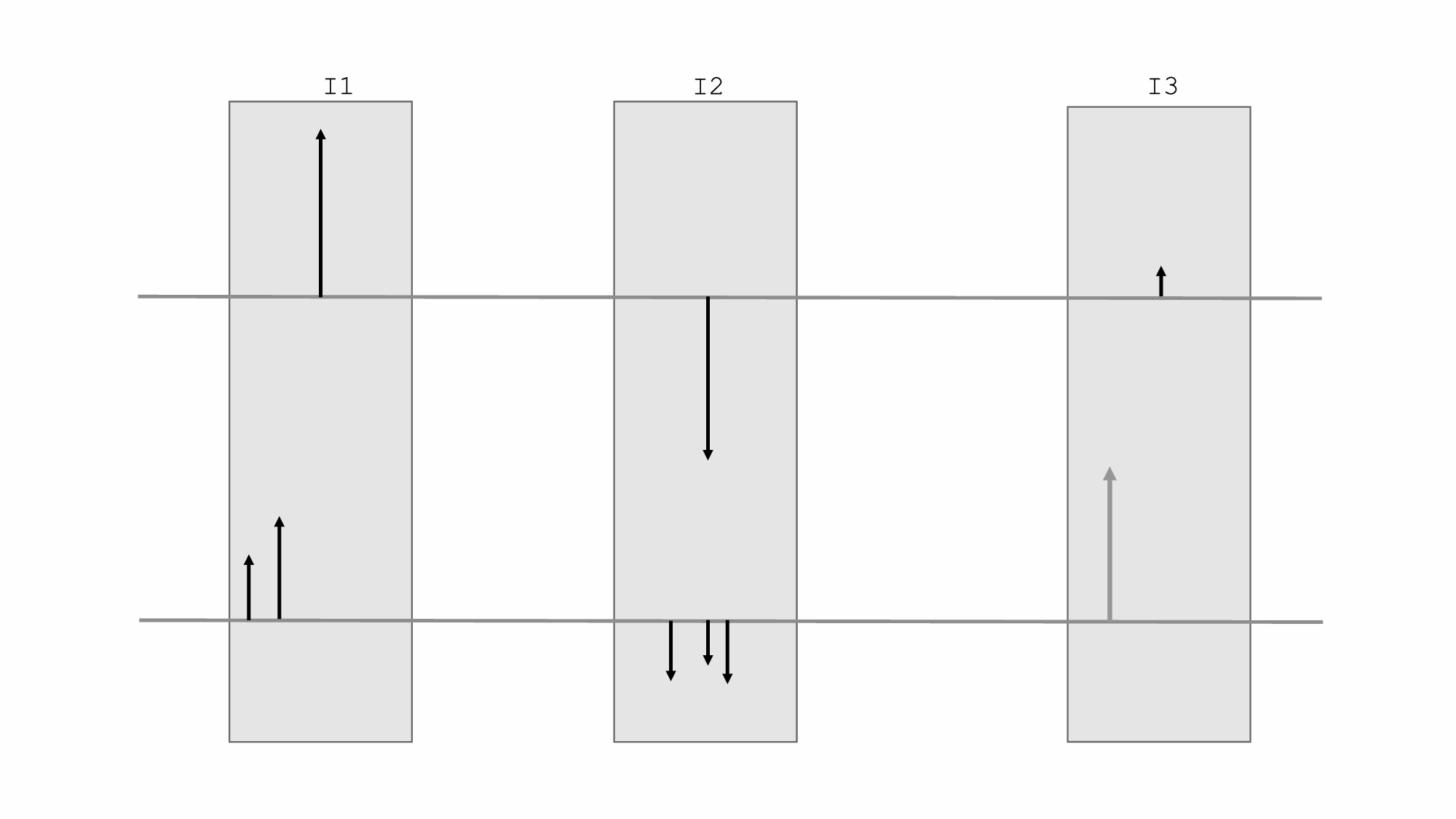}
\caption{}
\end{subfigure}
\begin{subfigure}[b]{0.4\textwidth}
\includegraphics[width=7cm]{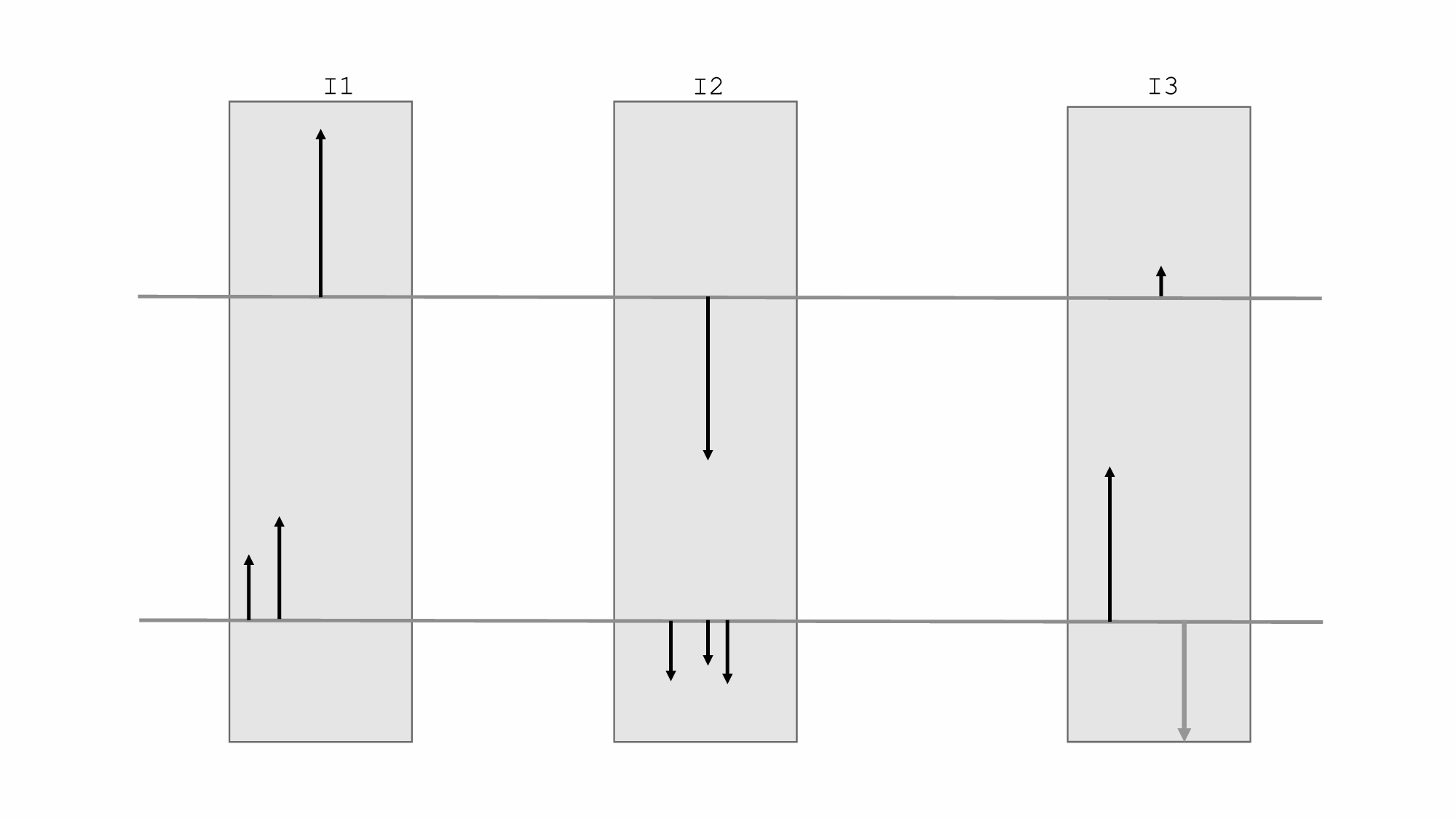}
\caption{}
\end{subfigure}
\caption{An illustration of the neighborhood concept: Two decompositions are shown, each by vertical arrows on a horizontal axes. The position of arrows shows $\theta$, while their amplitude shows $s$. The two decompositions are neighbor in a), while in b),c) and d) Conditions 1,2 and 3 are violated, respectively.}
\label{p2:fig:toplogy}
\end{figure}
Figure \ref{p2:fig:toplogy} illustrates the concept of $\epsilon-$neighborhood, where a decomposition is represented by a set of arrows, whose amplitudes show $s$, while their position denote $\theta$.  As seen, the definition does not restrict the orders. Condition 1 guarantees that the elements of $\{\bar{\theta}_k\}$ are concentrated around the elements of $\{\theta\}$. Then, Condition 2 provides that $\bar{A}$ leads to a close synthesis to $A$ through the model in \eqref{p2:eq:premodel}. Finally, Condition 3 guarantees that the LASSO cost values in \eqref{p2:eq:LASSO} for $A$ and $\bar{A}$ are close.

\label{p2:appendix:exist}

\section{Proof of Theorem \ref{p2:theorem:convergence}}
\label{p2:appendix:homotopyproof}
The proof is based on the following elements:
\begin{enumerate}
\item For a regular manifold, there exists a finite subset $\{\theta_{\text{b},1},\theta_{\text{b},1},\ldots,\theta_{\text{b},p}\}$, such that the matrix $\mathbf{B}=[\ba(\theta_{\text{b},1}),\ba(\theta_{\text{b},1}),\ldots,\ba(\theta_{\text{b},p})]$ is full rank.
\item The order of the LASSO and CLASS estimates are always bounded by $2mT$, i.e. the estimates are in $\calM_{2mT}$.
\item For any $R$ and $n$, the set $\calM_n^R$ of all decompositions with an order smaller than $n$ and amplitudes smaller than $R$, i.e. $|s_k(t)|<R$, is compact in the regular topology of fixed dimension. 
\item Define the synthesis function $f:\calM\to\mathbf{C}^{m\times T}$ such that for any $X=\{\bx(t)\}$ and $A=\{\theta_k,\{s_k(t)\}\}$ the relation $X=f(A)$ implies \eqref{p2:eq:premodel}. Also define $\ell(A)=\|\{s_k(t)\}\|_{2,1}$. Then $f$ and $\ell$ are continuous. 

\item For arbitrary observations $X=\{\bx(t)\}$, define also $\phi_{\tilde{\Theta}}(X)$ and $\phi(X)$ as the optimal values of the noiseless LASSO optimization in \eqref{p2:eq:LASSOnl} and the noiseless CLASS in \eqref{p2:eq:LASSOnlparametric}. Then, from the sparsity principle, we obtain that
\begin{eqnarray}\label{p2:eq:LASSOnlparametric1}
&\phi_{\tilde{\Theta}}(X)=\min\limits_{\tilde{\calM}}\|\{s_k(t)\}\|_{2,1}\nonumber\\
&\text{s.t}\nonumber\\
&\bx(t)=\sum\limits_{k=1}^n\as(\theta_k)s_k(t)
\end{eqnarray}
where the minimal point corresponds to the solution of LASSO \eqref{p2:eq:LASSOnl} .

\item The function $\phi_{\tilde{\Theta}}$ is convex and conic, i.e. for any observation sets $X,Y$ and $\alpha>0$, we have that $\phi_{\tilde{\Theta}}(X+Y)\leq\phi_{\tilde{\Theta}}(X)+\phi_{\tilde{\Theta}}(Y)$ and $\phi_{\tilde{\Theta}}(\alpha X)=\alpha\phi_{\tilde{\Theta}}(X)$.

\end{enumerate}
Using the above, the proof is straightforward. Note that from Observation 1, for any $X$ the solution $A$ to the group-LASSO optimization as well as the noiseless optimization is in $\calM_{2mT}^R$, where $R$ is a sufficiently large upper-bound on the amplitudes which only depends on $X$. If $X$ is further bounded, then $R$ is a constant.

Now, assume conversely that the theorem does not hold. This means that there exists an $\epsilon>0$ such that for any of the values $\delta_r=1/r$ there exists $X^{(r)}=\{\bx(t)+\bn^{(r)}(t)\}$, a $\delta_r-$dense grid $G_r$ and $\lambda_r<\delta_r$ such that $\|\bn^{(r)}(t)\|<\delta_r$, but their corresponding group-LASSO solution $A_r$ is out of the $\epsilon-$neghborhood of any solution $A$ to the noiseless CLASS for $X$. Since $X^{(r)}$ is bounded, there exists a fixed $R$, such that $A_r\in\calM_{2mT}^R$. Now, from the second observation, we may assume without loss of generality that the sequence $A_r$ has a limit $\bar{A}\in\calM_{2mT}^R$, since otherwise one may take a converging subsequence. But $\bar{A}$ is also out of the $\epsilon-$neighborhood of any solution $A$ of noiseless CLASS. We finally show in the sequel that in fact $\bar{A}$ is contrarily equal to a solution $A$, which completes the proof.

To show that $\bar{A}$ is a minimizer of noiseless CLASS, first
note that $\ba(\theta)$ is a continuous function over a compact set $\Theta$. Thus, it is uniformly continuous. This means that for each value $\mu>0$ there exists a $\delta>0$ such that $|\theta_1-\theta_2|<\delta$ implies that $\|\ba(\theta_1)-\ba(\theta_2)\|\leq\mu$. Fix a $\mu$ and corresponding $\delta$. Consider the noiseless CLASS solution $A=\{(\theta_k,\{s_k(t)\})\}$ of $X$. As $G_r$ is $\delta_r-$dense, for each $\theta_k$, there exists a $\htheta_k\in G_r$ such that $|\htheta_k-\theta_k|<\delta_r$. For sufficiently large $r$, this implies that $|\htheta_k-\theta_k|<\delta_r<\delta$, which further implies that $\|\ba(\htheta_k)-\ba(\theta_k)\|<\mu$. Take the approximate on-grid estimate $\hat{A}_r=\{(\htheta_k,\{s_k(t)\})\}\in\calM_{G_r}$ and define $\hat{X}_r=\{\hat{\bx}^{(r)}(t)\}=f(\hat{A}_r)$. Note that
\begin{equation}\label{p2:eq:3mid}
\phi_r(\hat{X}_r)\leq\ell(\hat{A}_r)=\ell(A)=\phi(X)
\end{equation} 
where $\phi_r=\phi_{G_r}$ and the right-hand side of the inequality is the cost calculated at $\hat{A}_r$. On the other hand, for large $r$
\begin{equation}
\|\hat{\bx}^{(r)}(t)-\bx(t)\|=\left\|\sum\limits_{k=1}^n(\as(\theta_k)-\as(\htheta_k))s_k(t)\right\|\leq\mu\sum\limits_{k=1}^n|s_k(t)|\leq\mu L
\end{equation}
where $L$ is a proper upper bound for $\sum\limits_{k=1}^n|s_k(t)|$ over time. This shows that
\begin{equation}
\lim_{r\to\infty}\hat{\bx}^{(r)}(t)=\bx(t)
\end{equation}

Note also that the group-LASSO in the parametric form \eqref{p2:eq:LASSOparametric1} can also be written as
\begin{eqnarray}\label{p2:eq:LASSOparametric2}
&\min\limits_{\calM_G,\{\by(t)\}}\frac{1}{2}\sum\limits_{t=1}^T\left\|\bx(t)-\by(t)\right\|_2^2
+\lambda\|\{s_k(t)\}\|_{2,1}\nonumber\\
&\text{s.t}\nonumber\\
&\by(t)=\sum\limits_{k=1}^n\as(\theta_k)s_k(t)
\end{eqnarray}
which can be simplified to
\begin{eqnarray}\label{p2:eq:LASSOparametric3}
\min\limits_{Y=\{\by(t)\}}\frac{1}{2}\sum\limits_{t=1}^T\left\|\bx(t)-\by(t)\right\|_2^2
+\lambda\phi_G(Y)
\end{eqnarray}

Consider, $X=X^{(r)}$, $\lambda=\lambda_r$ and $G=G_r$ . Then, the optimal point in \eqref{p2:eq:LASSOparametric2} and \eqref{p2:eq:LASSOparametric3} is given by $A_r$ and $Y^{(r)}=\{\by^{(r)}(t)\}=f(A_r)=f(A_r)$ respectively. Then,
\begin{equation}
\frac{1}{2}\sum\limits_{t=1}^T\left\|\bx(t)+\bn^{(r)}(t)-\by^{(r)}(t)\right\|_2^2
+\lambda_r\phi_r(Y^{(r)})\leq\frac{1}{2}\sum\limits_{t=1}^T\left\|\bx(t)+\bn^{(r)}(t)-\hat{\bx}^{(r)}(t)\right\|_2^2
+\lambda_r\phi_r(\hat{X}^{(r)})
\end{equation}
The right hand side is the cost in \eqref{p2:eq:LASSOparametric3} evaluated at $\hat{X}^{(r)}$. Then, using  \eqref{p2:eq:3mid}, we obtain that
\begin{equation}
\frac{1}{2}\sum\limits_{t=1}^T\left\|\bx(t)-\by^{(r)}(t)\right\|_2^2\leq
\frac{1}{2}\sum\limits_{t=1}^T\left\|\bx(t)-\hat{\bx}^{(r)}(t)\right\|_2^2+T\delta_r^2+\lambda_r\phi(X)
\end{equation}
Letting $r$ tend to infinity, we get that
\begin{equation}
\lim_{r\to\infty}\by^{(r)}(t)=\bx(t)
\end{equation}
Until now, we have found observations $Y^{r}$ converging to $X$ such that $A_r$ is the noiseless-LASSO solutions of $Y^r$ over $G_r$, i.e $\ell(A_r)=\phi_r(Y_r)$. Note that from the continuity of $f$ and $\ell$ we have that $f(\bar{A})=X$ and
\begin{equation}\label{p2:eq:4mid}
\ell(\bar{A})=\lim_{r\to\infty}\ell(A_r)
\end{equation} 

Define $E_r=\{\be^{(r)}(t)=\by^{(r)}(t)-\hat{\bx}^{(r)}(t)\}$. Then, $E_r$ tends to zero as $r$ tends to infinity. Then from observation 6,
\begin{equation}\label{p2:eq:5mid}
\ell(A_r)=\phi_r(Y_r)\leq\phi_r(\hat{X}_r)+\phi_r(E_r)\leq\ell(\hat(A)_r)+\phi_r(E_r)=\ell(A)+\phi_r(E_r)
\end{equation} 
The final observation is that $\phi_r(E_r)$ vanishes as $r$ tends to infinity. To see this consider the set in observation 1 and note that for an arbitrary $\mu$ and a large $r$ there exists indexes $\htheta^{r}_{\text{b},k}$ such that $\|\ba(\htheta{r}_{\text{b},k})-\ba(\theta_{\text{b},k})\|<\mu$. Define $\hat{\mathbf{B}}^r=[\ba(\htheta^r_{\text{b},1})\ldots\ba(\htheta^r_{\text{b},p})]$. As the set of full-rank matrices is open, $\mu$ can be selected such that $\hat{\mathbf{B}}^r$ is full rank. Then,
\begin{equation}
\phi_r(E_r)\leq\sum\limits_t\|(\hat{\mathbf{B}}^r)^\dagger\mathbf{e}^{(r)}(t)\|_2
\end{equation}  
which tends to zero as $\mathbf{e}^{(r)}$ vanishes and the pseudo inverse $(\hat{\mathbf{B}}^r)^\dagger$ stays bounded in the $\mu-$neighborhood of $\mathbf{B}$. Finally as $\phi_r(E_r)$ tends to zero, taking the limit of \eqref{p2:eq:5mid} and combining with \eqref{p2:eq:4mid} we conclude that.
\begin{equation}
\ell(\bar{A})\leq\ell(\bar{A})
\end{equation}
which shows that $\bar{A}$ is a minimizer of the noiseless CLASS.

\section{Proof of Theorem \ref{p2:theorem:CLASSdual}}
\label{p2:appendix:convergence}
First, note that for a regular manifold $\ba(\theta)$, the set
\begin{equation}
L=\left\{\{\bz(t)\}\mid\forall\theta\quad \sum\limits_{t=1}^T|\ba^H(\theta)\bz(t)|^2\leq 1\right\}
\end{equation}
is compact. For any grid $G$, define
\begin{equation}
L_G=\left\{\{\bz(t)\}\mid\forall\theta\in G\quad \sum\limits_{t=1}^T|\ba^H(\theta)\bz(t)|^2\leq 1\right\}
\end{equation}
Then there exists a value $\delta$ such that for every $\delta-$dense grid $G$, the set $L_G$ is compact. Furthermore, for any value $\mu>0$, there exists a $\delta$ value such that  for every $\delta-$dense grid $G$,
\begin{equation}
L\subseteq L_G\subseteq L^\mu
\end{equation}
where
$L^\mu$ denotes the union of all closed $\mu-$neighborhoods of elements in $L$. Note also that $L^\mu$ is compact.

Now, consider an arbitrary solution $A=\{(\theta_k,\{s_k(t)\})\}$ of the noiseless CLASS. Take a sequence of $\delta_r=1/r-$dense grids $G_r$ such that $\theta_k\in G_r$ for all $k$ and $r$. Then, clearly $A\in\calM_{G_r}$ and thus it minimizes noiseless group-LASSO over $G_r$. From Theorem \ref{p2:fact:KKT}, this means that there exists a sequence of dual vectors $Z_r=\{\bz_r(t)\}\in L_{G_r}$ such that
\begin{equation}\label{p2:eq:nlcond}
\ba^H(\theta_k)\bz_r(t)=\frac{s_k(t)}{p_k}
\end{equation}
Note that for any fixed $\mu$ and sufficiently large $r$ we have that $Z_r\in L^\mu$. Thus, $Z_r$ has a subsequence converging to a point $Z\in L^\mu$. Then
\begin{equation}
Z\in\bigcap\limits_\mu L^\mu=L
\end{equation}
since the choice of $\mu$ is arbitrary. clearly $Z$ also satisfies the other condition in \eqref{p2:eq:nlcond}.

Conversely, suppose that there exists  $Z_r=\{\bz(t)\}\in L$ satisfying  \eqref{p2:eq:nlcond} for $A$. Then, we show that $A$ is the global minimum of noiseless-CLASS. Take any other decomposition $B=\{(\theta^\prime_l,\{s^\prime_l(t)\})\}$ with $f(B)=f(A)=X$. Take the grid $G=\{\theta^\prime_l\}\cup\{\theta_k\}$. Note that taking the dual verifiers in $Z$, the conditions of Theorem \ref{p2:fact:KKT} for the noiseless case is satisfied. Thus, $A$ a minimizer of noiseless LASSO for grid $G$ and input $X$, which implies that $\ell(A)\leq\ell(B)$.

Finally, let us prove convergence. Suppose conversely that taking $\delta_r=1/r$, there exists a sequence of primal solutions $A_r$ with corresponding dual parameters $Z_r$ to the group lasso with a perturbed input $\{\bx(t)+\bn_r(t)\}$ where $\|\bn_r(t)\|_2\leq\delta_r$, $\lambda_r<\delta_r$ and over the $\delta_r-$dense grid, such that  $Z_r$ is not in a $\epsilon-$neighborhood of any dual vector of the noiseless CLASS solution. But since $Z_r$ can be bounded in a compact set for large enough $r$ and due to Theorem \ref{p2:theorem:convergence}, the sequences has a subsequence converging to $A$ and $Z$ respectively. Since $Z_r$ is $\epsilon-$distant from any dual solution of noiseless CLASS, the limit is so. But, it is simple to check that the conditions of the current theorem holds for $Z$, which implies that $Z$ is a dual for $A$. This shows contradiction and completes the proof.

\section{Proof of Theorem \ref{p2:theorem:pert}}
\label{p2:appendix:linearization}
First, let us explain part (a) with more details. Convergence means that:

For any $\omega>0$ there exists a $\delta>0$ such that if the grid is $\delta-$dense, $\lambda<\delta$ and perturbations satisfy $\|\bn(t)\|<\delta$ and the solution $A$ is in $\epsilon-$neighborhood of a noiseless solution $A_0$ such that $\epsilon<\delta$ then, the $\epsilon-$false alarms are in $\omega\delta-$neighborhood of $\bar{A}$ and $|\pi_l-\bar{\pi}_l|<\omega\delta$ and $|\sigma_l(t)-\bar{\sigma}_l(t)|<\omega\delta$ hold. 

Now, to prove this, we follow the following steps:

\begin{enumerate}
\item Suppose that $G=(\bpi,\bsigma,\bar{A})$ minimizes $g$ for a certain choice of $\bn(t)$, and true parameters and $H=(\bpi^\prime,\bsigma^\prime,  \bar{A}^\prime)$ is another non-optimal point. Then, there exists a constant $K$ depending only on true parameters such that 
\begin{equation}\label{p2:eq:append:mid0}
g(\bpi^\prime,\bsigma^\prime,\bar{A}^\prime)-g(\bpi,\bsigma,\bar{A})\geq (G,H)^2
\end{equation}
\item Consider $\epsilon$ and $\delta$ such that the solution of LASSO optimization with a $\delta-$dense grid $\tilde{\Theta}$, $\|\bn(t)\|<\delta$ and $\lambda<\delta$ is in $\epsilon$-neighborhood of the noiseless (true) solution. Denote by $\bpi_m,\bsigma_m,\bar{A}_m$ the corresponding parameters to the optimal point of LASSO with the optimal cost $f_{\text{min}}$. Then,  
\begin{equation}\label{p2:eq:append:mid1}
|f_{\text{min}}-g(\bpi_m,\bsigma_m,\bar{A}_m)|<K_1\epsilon\|\bn\|\delta
\end{equation}  
\item Consider the same setup as above and remember that $\bpi,\bsigma$ and $\bar{A}$ minimize $g$. Take the optimization
\begin{eqnarray}
&\min\|\{\ts_k(t)\}\|_{1,2}\nonumber\\
&\text{s.t.}\nonumber\\
&\sigma_l(t)=\sum\limits_{k\mid|\ttheta^k-\theta_{0,l}|<\epsilon}\ts_k(t)-s_{l,0}(t)\nonumber\\
&\pi_l\gamma_l(t)=\sum\limits_{k\mid|\ttheta^k-\theta_{0,l}|<\epsilon}\ts_k(t)(\ttheta^k-\theta_{l,0})
\end{eqnarray}
and note that it has a solution $\{\ts_k(t)\}$ with only two active elements in each cloud. Take this solution and calcualte the original LASSO cost $f$ at this point. Then
\begin{equation}\label{p2:eq:append:mid2}
|f-g(\bpi,\bsigma,\bar{A})|<K_2\epsilon\|\bn\|\delta
\end{equation}

\item Putting \eqref{p2:eq:append:mid2} and \eqref{p2:eq:append:mid1} together, it is simple to conclude that
\begin{equation}\label{p2:eq:append:mid3}
g(\bpi_m,\bsigma_m,\bar{A}_m)-g(\bpi,\bsigma,\bar{A})<K_3\epsilon\delta^2
\end{equation}

\item Now, if (a) is not correct then there exists a $\omega$ such that for any arbitrary $\delta$ there exists a $\delta-$exact case such that $d(G,Gm)>\omega\delta$. Consider now that $\epsilon<K\omega^2/K_3$  we get from  \eqref{p2:eq:append:mid2} that
\begin{equation}
K\omega^2\delta^2<g(\bpi_m,\bsigma_m,\bar{A}_m)-g(\bpi,\bsigma,\bar{A})<K_3\epsilon\delta^2
\end{equation}

which leads to $K\omega^2/K_3\omega^2<\epsilon$ and contradicts to the choice of $\epsilon$. Thus, (a) holds.

\item Relations \eqref{p2:eq:append:mid2} and \eqref{p2:eq:append:mid1} imply
\begin{equation}\label{p2:eq:append:mid4}
f-f_{\text{min}}<K\epsilon\delta^2
\end{equation} 

\item Similar to step 1 if $\|\{\ts(t)-\ts_m(t)\}\|_\infty=d$, then one can conclude that
\begin{equation} \label{p2:eq:append:mid5}
f-f_{\text{min}}>K_4d^2
\end{equation}

\item Finally for any $\omega$ and sufficiently small $\delta$, the relation $\|\{\ts(t)-\ts_m(t)\}\|_\infty<\omega\delta$ must hold otherwise \eqref{p2:eq:append:mid4} and \eqref{p2:eq:append:mid5} will contradict again for small choice of $\epsilon$ and $\delta$. Then, $\delta_3<\|\{\ts(t)-\ts_m(t)\}\|_\infty$ proves the result.
 
\end{enumerate}

\section{Proof of Theorem \ref{p2:theorem:consistecy}}
\label{p2:appendix:consistecy}
For part (a), it is easy to plug \eqref{p2:eq:dualpure} in Theorem \ref{p2:theorem:CLASSdual} and check by direct calculation that \eqref{p2:eq:FAcond} ensures optimality of the true parameters,

For (b), since $\{\theta_k,\{s_k(t)\}\}$ is consistent the optimization
\begin{eqnarray}
&\min\limits_{\{\bz(t)\}}\sum\limits_t\|\bz(t)\|^2_2\nonumber\\
&\text{s.t.}\nonumber\\
&\sum\limits_t|\bz^H(t)\ba(\theta)|^2_2\leq 1 \quad
\ba^H(\theta_k)\bz(t)=\gamma_k(t)=\frac{s_k(t)}{\sqrt{\sum\limits_t|s_k(t)|^2}}
\end{eqnarray}
is feasible and has solution $\bz^\prime$. It is simple to see that from the KKT theorem $\bz^\prime$ can be written as
\begin{equation}
\bz^\prime=\sum\limits_l\ba(\theta_l^\prime)r_l\gamma^\prime_l(t)+
\sum\limits_k\ba(\theta_k)u_k
\end{equation}  
where $\{\theta^\prime_l\}$ is the set of all peaks of the spectrum $|\ba^H(\theta)\bz^\prime|$, thus including $\theta_k$, and $r_l,u_k$ are suitable dual parameters. This shows that $\bz^\prime(t)$ is in the range space of $\bA^\prime$ consisting of $\ba(\theta_l^\prime)$ as columns, i.e
\begin{equation}\label{p2:eq:thmid1}
\bz(t)=\bA^\prime\bsigma^\prime(t)
\end{equation}
Furthermore,
\begin{equation}\label{p2:eq:thmid2}
\ba^H(\theta^\prime_l){\bz^\prime}(t)=\gamma^\prime_l(t)\to{\bA^\prime}^H\bz(t)=\bgamma^\prime(t)
\end{equation}
and 
\begin{equation}\label{p2:eq:thmid3}
\frac{\partial\sum\limits_t|\bz^H(t)\ba(\theta)|^2_2}{\partial\theta}\mid_{\theta=\theta^\prime_l}=0
\to\sum\limits_t\Re(\gamma^\prime_l(t)\bd^H(\theta^\prime_l)\bz(t))=0
\end{equation} 
It is easy by direct calculation to show that \eqref{p2:eq:thmid1},\eqref{p2:eq:thmid2} and \eqref{p2:eq:thmid3} may only hold if $\bsigma^\prime$ is equal to $\bsigma_0$ in \eqref{p2:eq:pisigmanl} if $\bA$ and $\bgamma$ are replaced by their primed counterparts and the resulting $\bdelta$ is zero. Then, similar to part (a), the optimality condition directly leads to \eqref{p2:eq:FAcondnl} which establishes pure consistency for $\{\theta_l^\prime,\{s_l(t)=\gamma^\prime_l(t)\}\}$. 

\section{Proof of Theorem \ref{p2:theorem:statistics}}
\label{p2:appendix:statistics}

a) By definition, $\lambda_b$ can be written as
\begin{eqnarray}
\lambda_b &=\min\{\lambda\mid\forall\theta\ 
\sum\limits_t\left|\ba^H(\theta)\left(\bn(t)-\sum\limits_l(\ba_l\sigma_l(t)+\bd_l\gamma_l(t)\pi_l))\right)\right|^2\leq\lambda^2
\}\nonumber\\
&=\min\bigcap\limits_{\theta}\underbrace{
\{\lambda\mid
\sum\limits_t\left|\ba^H(\theta)\left(\bn(t)-\sum\limits_l(\ba_l\sigma_l(t)+\bd_l\gamma_l(t)\pi_l))\right)\right|^2\leq\lambda^2
\}
}_{S_\theta}
\end{eqnarray}
Note that the term $\ba_l\sigma_l(t)+\bd_l\gamma_l(t)\pi_l)$ is linear in $\lambda$. Thus, $S_\theta=\{\lambda\mid P_\theta(\lambda)\leq 0\}$ where $P_\theta(\lambda)$ is a quadratic function of $\lambda$. Note that if the case is purely consistent the leading term in $P_\theta$ can be shown by calculation to be negative. Furthermore $P_\theta(0)>0$. Thus, $P_\theta$ has exactly one positive root $\Lambda(\theta)$, given by \eqref{p2:eq:Lambda}, and $S_\theta=[\Lambda(\theta)\ \infty)$, leading to
\begin{equation}
\lambda_b=\min\bigcap\limits_\theta[\Lambda(\theta)\ \infty)=\min[\max\limits_\theta\Lambda(\theta)\ \infty)=\max\limits_\theta\Lambda(\theta)
\end{equation}

b) The result follows from direct calculation and noting that $\calE(n(t)\lambda_b(n(t)))=0$. To see this follow the following steps

\begin{enumerate}
\item Note that $\lambda_b=\lambda_b(\{\bn(t)\})$ is conic function of noise, i.e. $\lambda_b(\{\alpha\bn(t)\})=|\alpha|\lambda_b(\{\bn(t)\})$.
\item Then,
\begin{equation}
\calE(\bn(t)\mid\lambda_b)=\lambda_b\calE(\bn(t)\mid\lambda_b=1)
\end{equation}
\item Note that $\calE(\bn(t)\mid\lambda_b=1)=0$, since
\begin{equation}
0=\calE(\bn(t))=\calE_{\lambda_b}(\calE(\bn(t)\mid\lambda_b))=
\calE(\bn(t)\mid\lambda_b=1)\calE(\lambda_b)
\end{equation} 

\item Finally,
\begin{equation}
\calE(n(t)\lambda_b)=\calE_{\lambda_b}(\lambda_b\calE(\bn(t)\mid\lambda_b))=
\calE(\bn(t)\mid\lambda_b=1)\calE(\lambda_b^2)=0
\end{equation}
\end{enumerate}

\ifCLASSOPTIONcaptionsoff
  \newpage
\fi
\bibliographystyle{IEEEtran}
\bibliography{IEEEabrv,bibFile}

\begin{thebibliography}{10}
\providecommand{\url}[1]{#1}
\csname url@samestyle\endcsname
\providecommand{\newblock}{\relax}
\providecommand{\bibinfo}[2]{#2}
\providecommand{\BIBentrySTDinterwordspacing}{\spaceskip=0pt\relax}
\providecommand{\BIBentryALTinterwordstretchfactor}{4}
\providecommand{\BIBentryALTinterwordspacing}{\spaceskip=\fontdimen2\font plus
\BIBentryALTinterwordstretchfactor\fontdimen3\font minus
  \fontdimen4\font\relax}
\providecommand{\BIBforeignlanguage}[2]{{%
\expandafter\ifx\csname l@#1\endcsname\relax
\typeout{** WARNING: IEEEtran.bst: No hyphenation pattern has been}%
\typeout{** loaded for the language `#1'. Using the pattern for}%
\typeout{** the default language instead.}%
\else
\language=\csname l@#1\endcsname
\fi
#2}}
\providecommand{\BIBdecl}{\relax}
\BIBdecl
\renewcommand{\BIBentryALTinterwordstretchfactor}{4}

\bibitem{genomics}
T.~T.˜Wu, Y.~F.˜Chen, T.˜Hastie, E.˜Sobel, and K.˜Lange, ``{Genome-wide
  association analysis by lasso penalized logistic regression},''
  \emph{Bioinformatics}, vol.~25, pp. 714--721, Mar. 2009.

\bibitem{parvaresh2008recovering}
F.˜Parvaresh, H.˜Vikalo, S.˜Misra, and B.˜Hassibi, ``Recovering sparse signals
  using sparse measurement matrices in compressed dna microarrays,''
  \emph{IEEE, J. Select. Topics Signal Processing}, vol.~2, pp. 275--285, June
  2008.

\bibitem{EEG}
W.˜Tu and S.˜Sun, ``Spatial filter selection with lasso for eeg
  classification,'' in \emph{Advanced Data Mining and Applications}, Chongqing,
  China, 2010, pp. 142--149.

\bibitem{Finance}
H.˜Konno and H.˜Yamazaki, ``Mean-absolute deviation portfolio optimization
  model and its applications to tokyo stock market,'' \emph{Manage. Sci.},
  vol.~37, pp. 519--531, May 1991.

\bibitem{yao2011compressive}
H.˜Yao, P.˜Gerstoft, P.~M.˜Shearer, and C.˜Mecklenbr{\"a}uker, ``Compressive
  sensing of the tohoku-oki mw 9.0 earthquake: Frequency-dependent rupture
  modes,'' \emph{Geophys. Res. Lett.}, vol.~38, Oct. 2011.

\bibitem{tibshirani}
R.˜Tibshirani, ``Regression shrinkage and selection via the lasso,'' \emph{J.
  Roy. Stat. Soc., Series B (Methodological)}, vol.~58, pp. 267--288, Jan.
  1996.

\bibitem{basis}
S.~S.˜Chen, D.~L.˜Donoho, and M.~A.˜Saunders, ``{Atomic Decomposition by Basis
  Pursuit},'' \emph{SIAM J. Sci. Comput.}, vol.~20, no.~1, pp. 33--61, Dec.
  1998.

\bibitem{fuchs_GMF}
J.~J.˜Fuchs, ``On the application of the global matched filter to doa
  estimation with uniform circular arrays,'' \emph{{IEEE} Trans. Signal
  Processing}, vol.~49, no.~4, pp. 702--709, Apr. 2001.

\bibitem{GPSR}
M.˜Figueiredo, R.˜Nowak, and S.˜Wright, ``Gradient projection for sparse
  reconstruction: Application to compressed sensing and other inverse
  problems,'' \emph{{IEEE} Select. Topic. Signal Processing}, vol.~1, pp.
  586--597, Dec. 2007.

\bibitem{AdaptiveLASSO}
H.˜Zou, ``The adaptive lasso and its oracle properties,'' \emph{J. Amer. Stat.
  Assoc.}, vol. 101, no. 476, pp. 1418--1429, 2006.

\bibitem{BLASSO}
T.˜Park and G.˜Casella, ``The bayesian lasso,'' \emph{J. Amer. Stat. Assoc.},
  vol. 103, pp. 681--686, 2008.

\bibitem{bayesianLASSO}
H.˜Zayyani, M.˜Babaie-Zadeh, and C.˜Jutten, ``Bayesian pursuit algorithm for
  sparse representation,'' in \emph{IEEE International Conference on Acoustics,
  Speech and Signal Processing, ICASSP}, Taipei, Taiwan, Apr. 2009, pp. 1549
  --1552.

\bibitem{MP}
S.~G.˜Mallat and Z.˜Zhang, ``Matching pursuits with time-frequency
  dictionaries,'' \emph{{IEEE} Trans. Signal Processing}, vol.~41, pp.
  3397--3415, Dec. 1993.

\bibitem{OMP}
J.~A.˜Tropp and A.~C.˜Gilbert, ``Signal recovery from random measurements via
  orthogonal matching pursuit,'' \emph{{IEEE} Trans. Inform. Theory}, vol.~53,
  pp. 4655--4666, Dec. 2007.

\bibitem{LARS}
B.˜Efron, T.˜Hastie, L.˜Johnstone, and R.˜Tibshirani, ``Least angle
  regression,'' \emph{Ann. Stat.}, vol.~32, pp. 407--499, Apr. 2004.

\bibitem{CoSaMP}
D.˜Needell and J.~A.˜Tropp, ``Cosamp: Iterative signal recovery from incomplete
  and inaccurate samples,'' \emph{Elsevier, Appl. Comput. Harmon. Anal.},
  vol.~26, pp. 301--321, May 2009.

\bibitem{lustig2007sparse}
M.˜Lustig, D.˜Donoho, and J.~M.˜Pauly, ``Sparse mri: The application of
  compressed sensing for rapid mr imaging,'' \emph{Resonance Med. Mag.},
  vol.~58, pp. 1182--1195, Dec. 2007.

\bibitem{provost2009application}
J.˜Provost and F.˜Lesage, ``The application of compressed sensing for
  photo-acoustic tomography,'' \emph{Medical Imaging, IEEE Transactions on},
  vol.~28, no.~4, pp. 585--594, 2009.

\bibitem{Eldar2009}
Y.~C.˜Eldar and M.˜Mishali, ``Robust recovery of signals from a structured
  union of subspaces,'' \emph{{IEEE} Trans. Inform. Theory}, vol.~55, no.~11,
  pp. 5302--5316, 2009.

\bibitem{Eldar2010}
M.˜Mishali and Y.~C.˜Eldar, ``From theory to practice: Sub-nyquist sampling of
  sparse wideband analog signals,'' \emph{IEEE J. Select. Topics Signal
  Processing}, vol.~4, pp. 375--391, Apr. 2010.

\bibitem{tropp2010beyond}
J.~A.˜Tropp, J.~N.˜Laska, M.~F.˜Duarte, J.~K.˜Romberg, and R.~G.˜Baraniuk,
  ``Beyond nyquist: Efficient sampling of sparse bandlimited signals,''
  \emph{{IEEE} Trans. Inform. Theory}, vol.~56, pp. 520--544, Jan. 2010.

\bibitem{Malioutov}
D.˜Malioutov, M.˜Cetin, and A.˜Willsky, ``Source localization by enforcing
  sparsity through a laplacian prior: an svd-based approach,'' \emph{{IEEE}
  Workshop Stat. Signal Processing}, pp. 573--576, Sept. 2003.

\bibitem{baraniuk2007compressive}
R.~G.˜Baraniuk, ``Compressive sensing [lecture notes],'' \emph{{IEEE} Signal
  Processing Mag.}, vol.~24, pp. 118--121, July 2007.

\bibitem{arigovindan2005variational}
M.˜Arigovindan, M.˜Suhling, P.˜Hunziker, and M.˜Unser, ``Variational image
  reconstruction from arbitrarily spaced samples: A fast multiresolution spline
  solution,'' \emph{{IEEE} Trans. Image Processing}, vol.~14, pp. 450--460,
  Apr. 2005.

\bibitem{milanfar1996moment}
P.˜Milanfar, W.~C.˜Karl, and A.~S.˜Willsky, ``A moment-based variational
  approach to tomographic reconstruction,'' \emph{{IEEE} Trans. Image
  Processing}, vol.~5, pp. 459--470, Mar. 1996.

\bibitem{herman2009high}
M.~A.˜Herman and T.˜Strohmer, ``High-resolution radar via compressed sensing,''
  \emph{Signal Processing, IEEE Transactions on}, vol.~57, no.~6, pp.
  2275--2284, 2009.

\bibitem{stoica1998matched}
P.˜Stoica, A.˜Jakobsson, and J.˜Li, ``Matched-filter bank interpretation of
  some spectral estimators,'' \emph{Signal Processing}, vol.~66, no.~1, pp.
  45--59, 1998.

\bibitem{zhao2006model}
P.˜Zhao and B.˜Yu, ``On model selection consistency of lasso,'' \emph{The
  Journal of Machine Learning Research}, vol.~7, pp. 2541--2563, 2006.

\bibitem{Donoho_CS}
D.˜Donoho, ``Compressed sensing,'' \emph{{IEEE} Trans. Inform. Theory},
  vol.~52, no.~4, pp. 1289 --1306, Apr. 2006.

\bibitem{candes-near}
E.~J.˜Cand{\`e}s and Y.˜Plan, ``Near-ideal model selection by $\ell_1$
  minimization,'' \emph{Ann. Stat.}, vol.~37, pp. 2145--2177, Oct. 2009.

\bibitem{RIPless}
E.~J.˜Candes and Y.˜Plan, ``{A Probabilistic and RIPless Theory of Compressed
  Sensing},'' \emph{{IEEE} Trans. Inform. Theory}, vol.~57, no.~11, pp.
  7235--7254, Nov. 2011.

\bibitem{oymak2013squared}
S.˜Oymak, C.˜Thrampoulidis, and B.˜Hassibi, ``The squared-error of generalized
  lasso: A precise analysis,'' \emph{arXiv preprint arXiv:1311.0830}, 2013.

\bibitem{candes2012towards}
E.˜Candes and C.˜Fernandez-Granda, ``Towards a mathematical theory of
  super-resolution,'' \emph{arXiv preprint arXiv:1203.5871}, 2012.

\bibitem{tang2012compressive}
G.˜Tang, B.~N.˜Bhaskar, P.˜Shah, and B.˜Recht, ``Compressive sensing off the
  grid,'' in \emph{Communication, Control, and Computing (Allerton), 2012 50th
  Annual Allerton Conference on}.\hskip 1em plus 0.5em minus 0.4em\relax IEEE,
  2012, pp. 778--785.

\bibitem{ben2009cramer}
Z.˜Ben-Haim and Y.~C.˜Eldar, ``The cramer-rao bound for sparse estimation,''
  \emph{arXiv preprint arXiv:0905.4378}, 2009.

\bibitem{stoica1990performance}
P.˜Stoica and A.˜Nehorai, ``Performance study of conditional and unconditional
  direction-of-arrival estimation,'' \emph{{IEEE} Trans. Acoust., Speech,
  Signal Processing}, vol.~38, no.~10, pp. 1783--1795, 1990.

\bibitem{bhaskar2011atomic}
B.~N.˜Bhaskar and B.˜Recht, ``Atomic norm denoising with applications to line
  spectral estimation,'' in \emph{Communication, Control, and Computing
  (Allerton), 2011 49th Annual Allerton Conference on}.\hskip 1em plus 0.5em
  minus 0.4em\relax IEEE, 2011, pp. 261--268.

\bibitem{Panahi2014Gridless}
M.~V.˜Ashkan~Panahi, ``Gridless compressive sensing,'' in \emph{{IEEE}Int.
  Conf. Acoust. Speech, Signal Processing}, 2014.

\bibitem{li1996efficient}
J.˜Li and P.˜Stoica, ``Efficient mixed-spectrum estimation with applications to
  target feature extraction,'' \emph{Signal Processing, IEEE Transactions on},
  vol.~44, no.~2, pp. 281--295, 1996.

\bibitem{fessler1994space}
J.~A.˜Fessler and A.~O.˜Hero, ``Space-alternating generalized
  expectation-maximization algorithm,'' \emph{Signal Processing, IEEE
  Transactions on}, vol.~42, no.~10, pp. 2664--2677, 1994.

\bibitem{donoho2003optimally}
D.~L.˜Donoho and M.˜Elad, ``Optimally sparse representation in general
  (nonorthogonal) dictionaries via l1 minimization,'' \emph{Proceedings of the
  National Academy of Sciences}, vol. 100, no.~5, pp. 2197--2202, 2003.

\bibitem{scharf1991statistical}
L.~L.˜Scharf, \emph{Statistical signal processing}.\hskip 1em plus 0.5em minus
  0.4em\relax Addison-Wesley Reading, MA, 1991, vol.~98.

\bibitem{stoica2004model}
P.˜Stoica and Y.˜Selen, ``Model-order selection: a review of information
  criterion rules,'' \emph{Signal Processing Magazine, IEEE}, vol.~21, no.~4,
  pp. 36--47, 2004.

\bibitem{rissanen1983universal}
J.˜Rissanen, ``A universal prior for integers and estimation by minimum
  description length,'' \emph{The Annals of statistics}, pp. 416--431, 1983.

\bibitem{stoica1989music}
P.˜Stoica and N.˜Arye, ``Music, maximum likelihood, and cramer-rao bound,''
  \emph{Acoustics, Speech and Signal Processing, IEEE Transactions on},
  vol.~37, no.~5, pp. 720--741, 1989.

\bibitem{ottersten1993exact}
B.˜Ottersten, M.˜Viberg, P.˜Stoica, and A.˜Nehorai, ``Exact and large sample
  {{ML}} techniques for parameter estimation and detection in array
  processing,'' in \emph{Radar Array Processing}, Haykin, Litva, and Shepherd,
  Eds.\hskip 1em plus 0.5em minus 0.4em\relax Berlin: Springer-Verlag, 1993,
  pp. 99--151.

\bibitem{Homotopy}
M.~R.˜Osborne, B.˜Presnell, and B.˜Turlach, ``A new approach to variable
  selection in least squares problems,'' 1999.

\bibitem{zhu2011sparsity}
H.˜Zhu, G.˜Leus, and G.~B.˜Giannakis, ``Sparsity-cognizant total least-squares
  for perturbed compressive sampling,'' \emph{Signal Processing, IEEE
  Transactions on}, vol.~59, no.~5, pp. 2002--2016, 2011.

\bibitem{Viberg_2decades}
H.˜Krim and M.˜Viberg, ``Two decades of array signal processing research: the
  parametric approach,'' \emph{{IEEE} Signal Processing Mag.}, vol.~13, pp. 67
  --94, July 1996.

\end{thebibliography}


\end{document}